\documentclass[longauth]{aaEC}  

\usepackage{natbib}
\usepackage{scalerel}
\usepackage{adjustbox}
\usepackage{txfonts}
\usepackage[utf8]{inputenc}
\usepackage[switch, modulo]{lineno}
\usepackage{balance}
\usepackage{graphicx}
\usepackage{txfonts}
\usepackage{multirow}
\usepackage{colortbl}
\usepackage{float}
\usepackage{soul}
\usepackage[table]{xcolor}

\usepackage{color}
\definecolor{blue}{rgb}{0., 0., 1}
\definecolor{lightblue}{rgb}{0.1,0.4,1.}
\usepackage[switch]{lineno} 

\newcommand {\T}{Table\,}
\newcommand {\Sec}{Sect.\,}
\newcommand {\Fig}{Fig.\,}
\newcommand {\Eq}{Eq.\,}
\newcommand {\HS}{\texttt{HST2EUCLID}}

\usepackage{hyperref}
\hypersetup{
    colorlinks=true,
    linkcolor=blue,
    filecolor=magenta,      
    urlcolor=blue,
    citecolor=blue
}

\makeatletter
\renewcommand*\aa@pageof{, page \thepage{} of \pageref*{LastPage}}
\makeatother

\usepackage{euclid}
\usepackage{orcidlink}
\let\orcid\orcidlink

\begin{document}


\title{\Euclid preparation}
\subtitle{LXXIV. Euclidised observations of Hubble Frontier Fields and CLASH galaxy clusters}    

\author{Euclid Collaboration: P.~Bergamini\orcid{0000-0003-1383-9414}\thanks{\email{pietro.bergamini@inaf.it}}\inst{\ref{aff1},\ref{aff2}}
\and M.~Meneghetti\orcid{0000-0003-1225-7084}\inst{\ref{aff2},\ref{aff3}}
\and G.~Angora\orcid{0000-0002-0316-6562}\inst{\ref{aff4},\ref{aff5}}
\and L.~Bazzanini\orcid{0000-0003-0727-0137}\inst{\ref{aff4},\ref{aff2}}
\and P.~Rosati\orcid{0000-0002-6813-0632}\inst{\ref{aff4},\ref{aff2}}
\and C.~Grillo\orcid{0000-0002-5926-7143}\inst{\ref{aff1},\ref{aff6}}
\and M.~Lombardi\orcid{0000-0002-3336-4965}\inst{\ref{aff1}}
\and D.~Abriola\inst{\ref{aff1}}
\and A.~Mercurio\orcid{0000-0001-9261-7849}\inst{\ref{aff5},\ref{aff7},\ref{aff8}}
\and F.~Calura\orcid{0000-0002-6175-0871}\inst{\ref{aff2}}
\and G.~Despali\orcid{0000-0001-6150-4112}\inst{\ref{aff9},\ref{aff2},\ref{aff3}}
\and J.~M.~Diego\orcid{0000-0001-9065-3926}\inst{\ref{aff10}}
\and R.~Gavazzi\orcid{0000-0002-5540-6935}\inst{\ref{aff11},\ref{aff12}}
\and P.~Hudelot\inst{\ref{aff12}}
\and L.~Leuzzi\inst{\ref{aff9},\ref{aff2}}
\and G.~Mahler\orcid{0000-0003-3266-2001}\inst{\ref{aff13},\ref{aff14},\ref{aff15}}
\and E.~Merlin\orcid{0000-0001-6870-8900}\inst{\ref{aff16}}
\and C.~Scarlata\orcid{0000-0002-9136-8876}\inst{\ref{aff17}}
\and N.~Aghanim\inst{\ref{aff18}}
\and B.~Altieri\orcid{0000-0003-3936-0284}\inst{\ref{aff19}}
\and A.~Amara\inst{\ref{aff20}}
\and S.~Andreon\orcid{0000-0002-2041-8784}\inst{\ref{aff21}}
\and N.~Auricchio\orcid{0000-0003-4444-8651}\inst{\ref{aff2}}
\and C.~Baccigalupi\orcid{0000-0002-8211-1630}\inst{\ref{aff22},\ref{aff23},\ref{aff24},\ref{aff25}}
\and M.~Baldi\orcid{0000-0003-4145-1943}\inst{\ref{aff26},\ref{aff2},\ref{aff3}}
\and S.~Bardelli\orcid{0000-0002-8900-0298}\inst{\ref{aff2}}
\and R.~Bender\orcid{0000-0001-7179-0626}\inst{\ref{aff27},\ref{aff28}}
\and C.~Bodendorf\inst{\ref{aff27}}
\and D.~Bonino\orcid{0000-0002-3336-9977}\inst{\ref{aff29}}
\and E.~Branchini\orcid{0000-0002-0808-6908}\inst{\ref{aff30},\ref{aff31},\ref{aff21}}
\and M.~Brescia\orcid{0000-0001-9506-5680}\inst{\ref{aff32},\ref{aff5},\ref{aff33}}
\and J.~Brinchmann\orcid{0000-0003-4359-8797}\inst{\ref{aff34},\ref{aff35}}
\and S.~Camera\orcid{0000-0003-3399-3574}\inst{\ref{aff36},\ref{aff37},\ref{aff29}}
\and V.~Capobianco\orcid{0000-0002-3309-7692}\inst{\ref{aff29}}
\and C.~Carbone\orcid{0000-0003-0125-3563}\inst{\ref{aff6}}
\and J.~Carretero\orcid{0000-0002-3130-0204}\inst{\ref{aff38},\ref{aff39}}
\and S.~Casas\orcid{0000-0002-4751-5138}\inst{\ref{aff40}}
\and F.~J.~Castander\orcid{0000-0001-7316-4573}\inst{\ref{aff41},\ref{aff42}}
\and M.~Castellano\orcid{0000-0001-9875-8263}\inst{\ref{aff16}}
\and G.~Castignani\orcid{0000-0001-6831-0687}\inst{\ref{aff2}}
\and S.~Cavuoti\orcid{0000-0002-3787-4196}\inst{\ref{aff5},\ref{aff33}}
\and A.~Cimatti\inst{\ref{aff43}}
\and C.~Colodro-Conde\inst{\ref{aff44}}
\and G.~Congedo\orcid{0000-0003-2508-0046}\inst{\ref{aff45}}
\and C.~J.~Conselice\orcid{0000-0003-1949-7638}\inst{\ref{aff46}}
\and L.~Conversi\orcid{0000-0002-6710-8476}\inst{\ref{aff47},\ref{aff19}}
\and Y.~Copin\orcid{0000-0002-5317-7518}\inst{\ref{aff48}}
\and F.~Courbin\orcid{0000-0003-0758-6510}\inst{\ref{aff49}}
\and H.~M.~Courtois\orcid{0000-0003-0509-1776}\inst{\ref{aff50}}
\and M.~Cropper\orcid{0000-0003-4571-9468}\inst{\ref{aff51}}
\and A.~Da~Silva\orcid{0000-0002-6385-1609}\inst{\ref{aff52},\ref{aff53}}
\and H.~Degaudenzi\orcid{0000-0002-5887-6799}\inst{\ref{aff54}}
\and G.~De~Lucia\orcid{0000-0002-6220-9104}\inst{\ref{aff23}}
\and A.~M.~Di~Giorgio\orcid{0000-0002-4767-2360}\inst{\ref{aff55}}
\and J.~Dinis\inst{\ref{aff52},\ref{aff53}}
\and M.~Douspis\inst{\ref{aff18}}
\and F.~Dubath\orcid{0000-0002-6533-2810}\inst{\ref{aff54}}
\and X.~Dupac\inst{\ref{aff19}}
\and S.~Dusini\orcid{0000-0002-1128-0664}\inst{\ref{aff56}}
\and M.~Farina\orcid{0000-0002-3089-7846}\inst{\ref{aff55}}
\and S.~Farrens\orcid{0000-0002-9594-9387}\inst{\ref{aff57}}
\and S.~Ferriol\inst{\ref{aff48}}
\and M.~Frailis\orcid{0000-0002-7400-2135}\inst{\ref{aff23}}
\and E.~Franceschi\orcid{0000-0002-0585-6591}\inst{\ref{aff2}}
\and M.~Fumana\orcid{0000-0001-6787-5950}\inst{\ref{aff6}}
\and S.~Galeotta\orcid{0000-0002-3748-5115}\inst{\ref{aff23}}
\and B.~Garilli\orcid{0000-0001-7455-8750}\thanks{Deceased}\inst{\ref{aff6}}
\and K.~George\orcid{0000-0002-1734-8455}\inst{\ref{aff28}}
\and B.~Gillis\orcid{0000-0002-4478-1270}\inst{\ref{aff45}}
\and C.~Giocoli\orcid{0000-0002-9590-7961}\inst{\ref{aff2},\ref{aff58}}
\and A.~Grazian\orcid{0000-0002-5688-0663}\inst{\ref{aff59}}
\and F.~Grupp\inst{\ref{aff27},\ref{aff28}}
\and L.~Guzzo\orcid{0000-0001-8264-5192}\inst{\ref{aff1},\ref{aff21}}
\and S.~V.~H.~Haugan\orcid{0000-0001-9648-7260}\inst{\ref{aff60}}
\and W.~Holmes\inst{\ref{aff61}}
\and I.~Hook\orcid{0000-0002-2960-978X}\inst{\ref{aff62}}
\and F.~Hormuth\inst{\ref{aff63}}
\and A.~Hornstrup\orcid{0000-0002-3363-0936}\inst{\ref{aff64},\ref{aff65}}
\and K.~Jahnke\orcid{0000-0003-3804-2137}\inst{\ref{aff66}}
\and E.~Keih\"anen\orcid{0000-0003-1804-7715}\inst{\ref{aff67}}
\and S.~Kermiche\orcid{0000-0002-0302-5735}\inst{\ref{aff68}}
\and A.~Kiessling\orcid{0000-0002-2590-1273}\inst{\ref{aff61}}
\and M.~Kilbinger\orcid{0000-0001-9513-7138}\inst{\ref{aff57}}
\and B.~Kubik\orcid{0009-0006-5823-4880}\inst{\ref{aff48}}
\and M.~K\"ummel\orcid{0000-0003-2791-2117}\inst{\ref{aff28}}
\and M.~Kunz\orcid{0000-0002-3052-7394}\inst{\ref{aff69}}
\and H.~Kurki-Suonio\orcid{0000-0002-4618-3063}\inst{\ref{aff70},\ref{aff71}}
\and R.~Laureijs\inst{\ref{aff72}}
\and S.~Ligori\orcid{0000-0003-4172-4606}\inst{\ref{aff29}}
\and P.~B.~Lilje\orcid{0000-0003-4324-7794}\inst{\ref{aff60}}
\and V.~Lindholm\orcid{0000-0003-2317-5471}\inst{\ref{aff70},\ref{aff71}}
\and I.~Lloro\inst{\ref{aff73}}
\and G.~Mainetti\orcid{0000-0003-2384-2377}\inst{\ref{aff74}}
\and D.~Maino\inst{\ref{aff1},\ref{aff6},\ref{aff75}}
\and E.~Maiorano\orcid{0000-0003-2593-4355}\inst{\ref{aff2}}
\and O.~Mansutti\orcid{0000-0001-5758-4658}\inst{\ref{aff23}}
\and O.~Marggraf\orcid{0000-0001-7242-3852}\inst{\ref{aff76}}
\and K.~Markovic\orcid{0000-0001-6764-073X}\inst{\ref{aff61}}
\and M.~Martinelli\orcid{0000-0002-6943-7732}\inst{\ref{aff16},\ref{aff77}}
\and N.~Martinet\orcid{0000-0003-2786-7790}\inst{\ref{aff11}}
\and F.~Marulli\orcid{0000-0002-8850-0303}\inst{\ref{aff9},\ref{aff2},\ref{aff3}}
\and R.~Massey\orcid{0000-0002-6085-3780}\inst{\ref{aff15}}
\and S.~Maurogordato\inst{\ref{aff78}}
\and E.~Medinaceli\orcid{0000-0002-4040-7783}\inst{\ref{aff2}}
\and S.~Mei\orcid{0000-0002-2849-559X}\inst{\ref{aff79}}
\and Y.~Mellier\inst{\ref{aff80},\ref{aff12}}
\and G.~Meylan\inst{\ref{aff49}}
\and M.~Moresco\orcid{0000-0002-7616-7136}\inst{\ref{aff9},\ref{aff2}}
\and L.~Moscardini\orcid{0000-0002-3473-6716}\inst{\ref{aff9},\ref{aff2},\ref{aff3}}
\and E.~Munari\orcid{0000-0002-1751-5946}\inst{\ref{aff23},\ref{aff22}}
\and R.~Nakajima\inst{\ref{aff76}}
\and C.~Neissner\orcid{0000-0001-8524-4968}\inst{\ref{aff81},\ref{aff39}}
\and S.-M.~Niemi\inst{\ref{aff72}}
\and J.~W.~Nightingale\orcid{0000-0002-8987-7401}\inst{\ref{aff82},\ref{aff15}}
\and C.~Padilla\orcid{0000-0001-7951-0166}\inst{\ref{aff81}}
\and S.~Paltani\orcid{0000-0002-8108-9179}\inst{\ref{aff54}}
\and F.~Pasian\orcid{0000-0002-4869-3227}\inst{\ref{aff23}}
\and K.~Pedersen\inst{\ref{aff83}}
\and V.~Pettorino\inst{\ref{aff72}}
\and S.~Pires\orcid{0000-0002-0249-2104}\inst{\ref{aff57}}
\and G.~Polenta\orcid{0000-0003-4067-9196}\inst{\ref{aff84}}
\and M.~Poncet\inst{\ref{aff85}}
\and L.~A.~Popa\inst{\ref{aff86}}
\and L.~Pozzetti\orcid{0000-0001-7085-0412}\inst{\ref{aff2}}
\and F.~Raison\orcid{0000-0002-7819-6918}\inst{\ref{aff27}}
\and R.~Rebolo\inst{\ref{aff44},\ref{aff87}}
\and A.~Renzi\orcid{0000-0001-9856-1970}\inst{\ref{aff88},\ref{aff56}}
\and J.~Rhodes\inst{\ref{aff61}}
\and G.~Riccio\inst{\ref{aff5}}
\and E.~Romelli\orcid{0000-0003-3069-9222}\inst{\ref{aff23}}
\and M.~Roncarelli\orcid{0000-0001-9587-7822}\inst{\ref{aff2}}
\and E.~Rossetti\orcid{0000-0003-0238-4047}\inst{\ref{aff26}}
\and R.~Saglia\orcid{0000-0003-0378-7032}\inst{\ref{aff28},\ref{aff27}}
\and Z.~Sakr\orcid{0000-0002-4823-3757}\inst{\ref{aff89},\ref{aff90},\ref{aff91}}
\and A.~G.~S\'anchez\orcid{0000-0003-1198-831X}\inst{\ref{aff27}}
\and D.~Sapone\orcid{0000-0001-7089-4503}\inst{\ref{aff92}}
\and B.~Sartoris\orcid{0000-0003-1337-5269}\inst{\ref{aff28},\ref{aff23}}
\and M.~Schirmer\orcid{0000-0003-2568-9994}\inst{\ref{aff66}}
\and P.~Schneider\orcid{0000-0001-8561-2679}\inst{\ref{aff76}}
\and A.~Secroun\orcid{0000-0003-0505-3710}\inst{\ref{aff68}}
\and E.~Sefusatti\orcid{0000-0003-0473-1567}\inst{\ref{aff23},\ref{aff22},\ref{aff24}}
\and G.~Seidel\orcid{0000-0003-2907-353X}\inst{\ref{aff66}}
\and S.~Serrano\orcid{0000-0002-0211-2861}\inst{\ref{aff42},\ref{aff41},\ref{aff93}}
\and C.~Sirignano\orcid{0000-0002-0995-7146}\inst{\ref{aff88},\ref{aff56}}
\and G.~Sirri\orcid{0000-0003-2626-2853}\inst{\ref{aff3}}
\and L.~Stanco\orcid{0000-0002-9706-5104}\inst{\ref{aff56}}
\and J.~Steinwagner\inst{\ref{aff27}}
\and P.~Tallada-Cresp\'{i}\orcid{0000-0002-1336-8328}\inst{\ref{aff38},\ref{aff39}}
\and H.~I.~Teplitz\orcid{0000-0002-7064-5424}\inst{\ref{aff94}}
\and I.~Tereno\inst{\ref{aff52},\ref{aff95}}
\and R.~Toledo-Moreo\orcid{0000-0002-2997-4859}\inst{\ref{aff96}}
\and F.~Torradeflot\orcid{0000-0003-1160-1517}\inst{\ref{aff39},\ref{aff38}}
\and I.~Tutusaus\orcid{0000-0002-3199-0399}\inst{\ref{aff90}}
\and L.~Valenziano\orcid{0000-0002-1170-0104}\inst{\ref{aff2},\ref{aff97}}
\and T.~Vassallo\orcid{0000-0001-6512-6358}\inst{\ref{aff28},\ref{aff23}}
\and G.~Verdoes~Kleijn\orcid{0000-0001-5803-2580}\inst{\ref{aff98}}
\and A.~Veropalumbo\orcid{0000-0003-2387-1194}\inst{\ref{aff21},\ref{aff31},\ref{aff99}}
\and Y.~Wang\orcid{0000-0002-4749-2984}\inst{\ref{aff94}}
\and J.~Weller\orcid{0000-0002-8282-2010}\inst{\ref{aff28},\ref{aff27}}
\and G.~Zamorani\orcid{0000-0002-2318-301X}\inst{\ref{aff2}}
\and E.~Zucca\orcid{0000-0002-5845-8132}\inst{\ref{aff2}}
\and A.~Biviano\orcid{0000-0002-0857-0732}\inst{\ref{aff23},\ref{aff22}}
\and M.~Bolzonella\orcid{0000-0003-3278-4607}\inst{\ref{aff2}}
\and A.~Boucaud\orcid{0000-0001-7387-2633}\inst{\ref{aff79}}
\and E.~Bozzo\orcid{0000-0002-8201-1525}\inst{\ref{aff54}}
\and C.~Burigana\orcid{0000-0002-3005-5796}\inst{\ref{aff100},\ref{aff97}}
\and M.~Calabrese\orcid{0000-0002-2637-2422}\inst{\ref{aff101},\ref{aff6}}
\and D.~Di~Ferdinando\inst{\ref{aff3}}
\and J.~A.~Escartin~Vigo\inst{\ref{aff27}}
\and R.~Farinelli\inst{\ref{aff2}}
\and J.~Gracia-Carpio\inst{\ref{aff27}}
\and N.~Mauri\orcid{0000-0001-8196-1548}\inst{\ref{aff43},\ref{aff3}}
\and V.~Scottez\inst{\ref{aff80},\ref{aff102}}
\and M.~Tenti\orcid{0000-0002-4254-5901}\inst{\ref{aff3}}
\and M.~Viel\orcid{0000-0002-2642-5707}\inst{\ref{aff22},\ref{aff23},\ref{aff25},\ref{aff24},\ref{aff103}}
\and M.~Wiesmann\orcid{0009-0000-8199-5860}\inst{\ref{aff60}}
\and Y.~Akrami\orcid{0000-0002-2407-7956}\inst{\ref{aff104},\ref{aff105}}
\and V.~Allevato\orcid{0000-0001-7232-5152}\inst{\ref{aff5}}
\and S.~Anselmi\orcid{0000-0002-3579-9583}\inst{\ref{aff56},\ref{aff88},\ref{aff106}}
\and M.~Ballardini\orcid{0000-0003-4481-3559}\inst{\ref{aff4},\ref{aff2},\ref{aff107}}
\and M.~Bethermin\orcid{0000-0002-3915-2015}\inst{\ref{aff108},\ref{aff11}}
\and A.~Blanchard\orcid{0000-0001-8555-9003}\inst{\ref{aff90}}
\and L.~Blot\orcid{0000-0002-9622-7167}\inst{\ref{aff109},\ref{aff106}}
\and S.~Borgani\orcid{0000-0001-6151-6439}\inst{\ref{aff110},\ref{aff22},\ref{aff23},\ref{aff24}}
\and A.~S.~Borlaff\orcid{0000-0003-3249-4431}\inst{\ref{aff111},\ref{aff112}}
\and S.~Bruton\orcid{0000-0002-6503-5218}\inst{\ref{aff17}}
\and R.~Cabanac\orcid{0000-0001-6679-2600}\inst{\ref{aff90}}
\and A.~Calabro\orcid{0000-0003-2536-1614}\inst{\ref{aff16}}
\and G.~Ca\~nas-Herrera\orcid{0000-0003-2796-2149}\inst{\ref{aff72},\ref{aff113}}
\and A.~Cappi\inst{\ref{aff2},\ref{aff78}}
\and C.~S.~Carvalho\inst{\ref{aff95}}
\and T.~Castro\orcid{0000-0002-6292-3228}\inst{\ref{aff23},\ref{aff24},\ref{aff22},\ref{aff103}}
\and K.~C.~Chambers\orcid{0000-0001-6965-7789}\inst{\ref{aff114}}
\and S.~Contarini\orcid{0000-0002-9843-723X}\inst{\ref{aff27}}
\and T.~Contini\orcid{0000-0003-0275-938X}\inst{\ref{aff90}}
\and A.~R.~Cooray\orcid{0000-0002-3892-0190}\inst{\ref{aff115}}
\and O.~Cucciati\orcid{0000-0002-9336-7551}\inst{\ref{aff2}}
\and B.~De~Caro\inst{\ref{aff6}}
\and G.~Desprez\inst{\ref{aff116}}
\and A.~D\'iaz-S\'anchez\orcid{0000-0003-0748-4768}\inst{\ref{aff117}}
\and S.~Di~Domizio\orcid{0000-0003-2863-5895}\inst{\ref{aff30},\ref{aff31}}
\and H.~Dole\orcid{0000-0002-9767-3839}\inst{\ref{aff18}}
\and S.~Escoffier\orcid{0000-0002-2847-7498}\inst{\ref{aff68}}
\and A.~G.~Ferrari\orcid{0009-0005-5266-4110}\inst{\ref{aff43},\ref{aff3}}
\and I.~Ferrero\orcid{0000-0002-1295-1132}\inst{\ref{aff60}}
\and F.~Finelli\orcid{0000-0002-6694-3269}\inst{\ref{aff2},\ref{aff97}}
\and F.~Fornari\orcid{0000-0003-2979-6738}\inst{\ref{aff97}}
\and L.~Gabarra\orcid{0000-0002-8486-8856}\inst{\ref{aff118}}
\and K.~Ganga\orcid{0000-0001-8159-8208}\inst{\ref{aff79}}
\and J.~Garc\'ia-Bellido\orcid{0000-0002-9370-8360}\inst{\ref{aff104}}
\and V.~Gautard\inst{\ref{aff119}}
\and E.~Gaztanaga\orcid{0000-0001-9632-0815}\inst{\ref{aff41},\ref{aff42},\ref{aff120}}
\and F.~Giacomini\orcid{0000-0002-3129-2814}\inst{\ref{aff3}}
\and G.~Gozaliasl\orcid{0000-0002-0236-919X}\inst{\ref{aff121},\ref{aff70}}
\and A.~Hall\orcid{0000-0002-3139-8651}\inst{\ref{aff45}}
\and H.~Hildebrandt\orcid{0000-0002-9814-3338}\inst{\ref{aff122}}
\and J.~Hjorth\orcid{0000-0002-4571-2306}\inst{\ref{aff123}}
\and M.~Huertas-Company\orcid{0000-0002-1416-8483}\inst{\ref{aff44},\ref{aff124},\ref{aff125},\ref{aff126}}
\and A.~Jimenez~Mu\~noz\orcid{0009-0004-5252-185X}\inst{\ref{aff127}}
\and J.~J.~E.~Kajava\orcid{0000-0002-3010-8333}\inst{\ref{aff128},\ref{aff129}}
\and V.~Kansal\orcid{0000-0002-4008-6078}\inst{\ref{aff130},\ref{aff131}}
\and D.~Karagiannis\orcid{0000-0002-4927-0816}\inst{\ref{aff132},\ref{aff133}}
\and C.~C.~Kirkpatrick\inst{\ref{aff67}}
\and L.~Legrand\orcid{0000-0003-0610-5252}\inst{\ref{aff134}}
\and G.~Libet\inst{\ref{aff85}}
\and A.~Loureiro\orcid{0000-0002-4371-0876}\inst{\ref{aff135},\ref{aff136}}
\and G.~Maggio\orcid{0000-0003-4020-4836}\inst{\ref{aff23}}
\and M.~Magliocchetti\orcid{0000-0001-9158-4838}\inst{\ref{aff55}}
\and C.~Mancini\orcid{0000-0002-4297-0561}\inst{\ref{aff6}}
\and F.~Mannucci\orcid{0000-0002-4803-2381}\inst{\ref{aff137}}
\and R.~Maoli\orcid{0000-0002-6065-3025}\inst{\ref{aff138},\ref{aff16}}
\and C.~J.~A.~P.~Martins\orcid{0000-0002-4886-9261}\inst{\ref{aff139},\ref{aff34}}
\and S.~Matthew\orcid{0000-0001-8448-1697}\inst{\ref{aff45}}
\and L.~Maurin\orcid{0000-0002-8406-0857}\inst{\ref{aff18}}
\and R.~B.~Metcalf\orcid{0000-0003-3167-2574}\inst{\ref{aff9},\ref{aff2}}
\and P.~Monaco\orcid{0000-0003-2083-7564}\inst{\ref{aff110},\ref{aff23},\ref{aff24},\ref{aff22}}
\and C.~Moretti\orcid{0000-0003-3314-8936}\inst{\ref{aff25},\ref{aff103},\ref{aff23},\ref{aff22},\ref{aff24}}
\and G.~Morgante\inst{\ref{aff2}}
\and Nicholas~A.~Walton\orcid{0000-0003-3983-8778}\inst{\ref{aff140}}
\and J.~Odier\orcid{0000-0002-1650-2246}\inst{\ref{aff127}}
\and L.~Patrizii\inst{\ref{aff3}}
\and A.~Pezzotta\orcid{0000-0003-0726-2268}\inst{\ref{aff27}}
\and M.~P\"ontinen\orcid{0000-0001-5442-2530}\inst{\ref{aff70}}
\and V.~Popa\inst{\ref{aff86}}
\and C.~Porciani\orcid{0000-0002-7797-2508}\inst{\ref{aff76}}
\and D.~Potter\orcid{0000-0002-0757-5195}\inst{\ref{aff141}}
\and I.~Risso\orcid{0000-0003-2525-7761}\inst{\ref{aff99}}
\and P.-F.~Rocci\inst{\ref{aff18}}
\and M.~Sahl\'en\orcid{0000-0003-0973-4804}\inst{\ref{aff142}}
\and A.~Schneider\orcid{0000-0001-7055-8104}\inst{\ref{aff141}}
\and M.~Sereno\orcid{0000-0003-0302-0325}\inst{\ref{aff2},\ref{aff3}}
\and P.~Simon\inst{\ref{aff76}}
\and A.~Spurio~Mancini\orcid{0000-0001-5698-0990}\inst{\ref{aff143},\ref{aff51}}
\and S.~A.~Stanford\orcid{0000-0003-0122-0841}\inst{\ref{aff144}}
\and C.~Tao\orcid{0000-0001-7961-8177}\inst{\ref{aff68}}
\and G.~Testera\inst{\ref{aff31}}
\and R.~Teyssier\orcid{0000-0001-7689-0933}\inst{\ref{aff145}}
\and S.~Toft\orcid{0000-0003-3631-7176}\inst{\ref{aff65},\ref{aff146},\ref{aff147}}
\and S.~Tosi\orcid{0000-0002-7275-9193}\inst{\ref{aff30},\ref{aff31}}
\and A.~Troja\orcid{0000-0003-0239-4595}\inst{\ref{aff88},\ref{aff56}}
\and M.~Tucci\inst{\ref{aff54}}
\and C.~Valieri\inst{\ref{aff3}}
\and J.~Valiviita\orcid{0000-0001-6225-3693}\inst{\ref{aff70},\ref{aff71}}
\and D.~Vergani\orcid{0000-0003-0898-2216}\inst{\ref{aff2}}
\and G.~Verza\orcid{0000-0002-1886-8348}\inst{\ref{aff148},\ref{aff149}}}
										   
\institute{Dipartimento di Fisica "Aldo Pontremoli", Universit\`a degli Studi di Milano, Via Celoria 16, 20133 Milano, Italy\label{aff1}
\and
INAF-Osservatorio di Astrofisica e Scienza dello Spazio di Bologna, Via Piero Gobetti 93/3, 40129 Bologna, Italy\label{aff2}
\and
INFN-Sezione di Bologna, Viale Berti Pichat 6/2, 40127 Bologna, Italy\label{aff3}
\and
Dipartimento di Fisica e Scienze della Terra, Universit\`a degli Studi di Ferrara, Via Giuseppe Saragat 1, 44122 Ferrara, Italy\label{aff4}
\and
INAF-Osservatorio Astronomico di Capodimonte, Via Moiariello 16, 80131 Napoli, Italy\label{aff5}
\and
INAF-IASF Milano, Via Alfonso Corti 12, 20133 Milano, Italy\label{aff6}
\and
Universita di Salerno, Dipartimento di Fisica "E.R. Caianiello", Via Giovanni Paolo II 132, I-84084 Fisciano (SA), Italy\label{aff7}
\and
INFN -- Gruppo Collegato di Salerno - Sezione di Napoli, Dipartimento di Fisica "E.R. Caianiello", Universita di Salerno, via Giovanni Paolo II, 132 - I-84084 Fisciano (SA), Italy\label{aff8}
\and
Dipartimento di Fisica e Astronomia "Augusto Righi" - Alma Mater Studiorum Universit\`a di Bologna, via Piero Gobetti 93/2, 40129 Bologna, Italy\label{aff9}
\and
Instituto de F\'isica de Cantabria, Edificio Juan Jord\'a, Avenida de los Castros, 39005 Santander, Spain\label{aff10}
\and
Aix-Marseille Universit\'e, CNRS, CNES, LAM, Marseille, France\label{aff11}
\and
Institut d'Astrophysique de Paris, UMR 7095, CNRS, and Sorbonne Universit\'e, 98 bis boulevard Arago, 75014 Paris, France\label{aff12}
\and
STAR Institute, Quartier Agora - All\'ee du six Ao\^ut, 19c B-4000 Li\`ege, Belgium\label{aff13}
\and
Department of Physics, Centre for Extragalactic Astronomy, Durham University, South Road, Durham, DH1 3LE, UK\label{aff14}
\and
Department of Physics, Institute for Computational Cosmology, Durham University, South Road, Durham, DH1 3LE, UK\label{aff15}
\and
INAF-Osservatorio Astronomico di Roma, Via Frascati 33, 00078 Monteporzio Catone, Italy\label{aff16}
\and
Minnesota Institute for Astrophysics, University of Minnesota, 116 Church St SE, Minneapolis, MN 55455, USA\label{aff17}
\and
Universit\'e Paris-Saclay, CNRS, Institut d'astrophysique spatiale, 91405, Orsay, France\label{aff18}
\and
ESAC/ESA, Camino Bajo del Castillo, s/n., Urb. Villafranca del Castillo, 28692 Villanueva de la Ca\~nada, Madrid, Spain\label{aff19}
\and
School of Mathematics and Physics, University of Surrey, Guildford, Surrey, GU2 7XH, UK\label{aff20}
\and
INAF-Osservatorio Astronomico di Brera, Via Brera 28, 20122 Milano, Italy\label{aff21}
\and
IFPU, Institute for Fundamental Physics of the Universe, via Beirut 2, 34151 Trieste, Italy\label{aff22}
\and
INAF-Osservatorio Astronomico di Trieste, Via G. B. Tiepolo 11, 34143 Trieste, Italy\label{aff23}
\and
INFN, Sezione di Trieste, Via Valerio 2, 34127 Trieste TS, Italy\label{aff24}
\and
SISSA, International School for Advanced Studies, Via Bonomea 265, 34136 Trieste TS, Italy\label{aff25}
\and
Dipartimento di Fisica e Astronomia, Universit\`a di Bologna, Via Gobetti 93/2, 40129 Bologna, Italy\label{aff26}
\and
Max Planck Institute for Extraterrestrial Physics, Giessenbachstr. 1, 85748 Garching, Germany\label{aff27}
\and
Universit\"ats-Sternwarte M\"unchen, Fakult\"at f\"ur Physik, Ludwig-Maximilians-Universit\"at M\"unchen, Scheinerstrasse 1, 81679 M\"unchen, Germany\label{aff28}
\and
INAF-Osservatorio Astrofisico di Torino, Via Osservatorio 20, 10025 Pino Torinese (TO), Italy\label{aff29}
\and
Dipartimento di Fisica, Universit\`a di Genova, Via Dodecaneso 33, 16146, Genova, Italy\label{aff30}
\and
INFN-Sezione di Genova, Via Dodecaneso 33, 16146, Genova, Italy\label{aff31}
\and
Department of Physics "E. Pancini", University Federico II, Via Cinthia 6, 80126, Napoli, Italy\label{aff32}
\and
INFN section of Naples, Via Cinthia 6, 80126, Napoli, Italy\label{aff33}
\and
Instituto de Astrof\'isica e Ci\^encias do Espa\c{c}o, Universidade do Porto, CAUP, Rua das Estrelas, PT4150-762 Porto, Portugal\label{aff34}
\and
Faculdade de Ci\^encias da Universidade do Porto, Rua do Campo de Alegre, 4150-007 Porto, Portugal\label{aff35}
\and
Dipartimento di Fisica, Universit\`a degli Studi di Torino, Via P. Giuria 1, 10125 Torino, Italy\label{aff36}
\and
INFN-Sezione di Torino, Via P. Giuria 1, 10125 Torino, Italy\label{aff37}
\and
Centro de Investigaciones Energ\'eticas, Medioambientales y Tecnol\'ogicas (CIEMAT), Avenida Complutense 40, 28040 Madrid, Spain\label{aff38}
\and
Port d'Informaci\'{o} Cient\'{i}fica, Campus UAB, C. Albareda s/n, 08193 Bellaterra (Barcelona), Spain\label{aff39}
\and
Institute for Theoretical Particle Physics and Cosmology (TTK), RWTH Aachen University, 52056 Aachen, Germany\label{aff40}
\and
Institute of Space Sciences (ICE, CSIC), Campus UAB, Carrer de Can Magrans, s/n, 08193 Barcelona, Spain\label{aff41}
\and
Institut d'Estudis Espacials de Catalunya (IEEC),  Edifici RDIT, Campus UPC, 08860 Castelldefels, Barcelona, Spain\label{aff42}
\and
Dipartimento di Fisica e Astronomia "Augusto Righi" - Alma Mater Studiorum Universit\`a di Bologna, Viale Berti Pichat 6/2, 40127 Bologna, Italy\label{aff43}
\and
Instituto de Astrof\'isica de Canarias, Calle V\'ia L\'actea s/n, 38204, San Crist\'obal de La Laguna, Tenerife, Spain\label{aff44}
\and
Institute for Astronomy, University of Edinburgh, Royal Observatory, Blackford Hill, Edinburgh EH9 3HJ, UK\label{aff45}
\and
Jodrell Bank Centre for Astrophysics, Department of Physics and Astronomy, University of Manchester, Oxford Road, Manchester M13 9PL, UK\label{aff46}
\and
European Space Agency/ESRIN, Largo Galileo Galilei 1, 00044 Frascati, Roma, Italy\label{aff47}
\and
Universit\'e Claude Bernard Lyon 1, CNRS/IN2P3, IP2I Lyon, UMR 5822, Villeurbanne, F-69100, France\label{aff48}
\and
Institute of Physics, Laboratory of Astrophysics, Ecole Polytechnique F\'ed\'erale de Lausanne (EPFL), Observatoire de Sauverny, 1290 Versoix, Switzerland\label{aff49}
\and
UCB Lyon 1, CNRS/IN2P3, IUF, IP2I Lyon, 4 rue Enrico Fermi, 69622 Villeurbanne, France\label{aff50}
\and
Mullard Space Science Laboratory, University College London, Holmbury St Mary, Dorking, Surrey RH5 6NT, UK\label{aff51}
\and
Departamento de F\'isica, Faculdade de Ci\^encias, Universidade de Lisboa, Edif\'icio C8, Campo Grande, PT1749-016 Lisboa, Portugal\label{aff52}
\and
Instituto de Astrof\'isica e Ci\^encias do Espa\c{c}o, Faculdade de Ci\^encias, Universidade de Lisboa, Campo Grande, 1749-016 Lisboa, Portugal\label{aff53}
\and
Department of Astronomy, University of Geneva, ch. d'Ecogia 16, 1290 Versoix, Switzerland\label{aff54}
\and
INAF-Istituto di Astrofisica e Planetologia Spaziali, via del Fosso del Cavaliere, 100, 00100 Roma, Italy\label{aff55}
\and
INFN-Padova, Via Marzolo 8, 35131 Padova, Italy\label{aff56}
\and
Universit\'e Paris-Saclay, Universit\'e Paris Cit\'e, CEA, CNRS, AIM, 91191, Gif-sur-Yvette, France\label{aff57}
\and
Istituto Nazionale di Fisica Nucleare, Sezione di Bologna, Via Irnerio 46, 40126 Bologna, Italy\label{aff58}
\and
INAF-Osservatorio Astronomico di Padova, Via dell'Osservatorio 5, 35122 Padova, Italy\label{aff59}
\and
Institute of Theoretical Astrophysics, University of Oslo, P.O. Box 1029 Blindern, 0315 Oslo, Norway\label{aff60}
\and
Jet Propulsion Laboratory, California Institute of Technology, 4800 Oak Grove Drive, Pasadena, CA, 91109, USA\label{aff61}
\and
Department of Physics, Lancaster University, Lancaster, LA1 4YB, UK\label{aff62}
\and
Felix Hormuth Engineering, Goethestr. 17, 69181 Leimen, Germany\label{aff63}
\and
Technical University of Denmark, Elektrovej 327, 2800 Kgs. Lyngby, Denmark\label{aff64}
\and
Cosmic Dawn Center (DAWN), Denmark\label{aff65}
\and
Max-Planck-Institut f\"ur Astronomie, K\"onigstuhl 17, 69117 Heidelberg, Germany\label{aff66}
\and
Department of Physics and Helsinki Institute of Physics, Gustaf H\"allstr\"omin katu 2, University of Helsinki, 00014 Helsinki, Finland\label{aff67}
\and
Aix-Marseille Universit\'e, CNRS/IN2P3, CPPM, Marseille, France\label{aff68}
\and
Universit\'e de Gen\`eve, D\'epartement de Physique Th\'eorique and Centre for Astroparticle Physics, 24 quai Ernest-Ansermet, CH-1211 Gen\`eve 4, Switzerland\label{aff69}
\and
Department of Physics, P.O. Box 64, University of Helsinki, 00014 Helsinki, Finland\label{aff70}
\and
Helsinki Institute of Physics, Gustaf H{\"a}llstr{\"o}min katu 2, University of Helsinki, Helsinki, Finland\label{aff71}
\and
European Space Agency/ESTEC, Keplerlaan 1, 2201 AZ Noordwijk, The Netherlands\label{aff72}
\and
NOVA optical infrared instrumentation group at ASTRON, Oude Hoogeveensedijk 4, 7991PD, Dwingeloo, The Netherlands\label{aff73}
\and
Centre de Calcul de l'IN2P3/CNRS, 21 avenue Pierre de Coubertin 69627 Villeurbanne Cedex, France\label{aff74}
\and
INFN-Sezione di Milano, Via Celoria 16, 20133 Milano, Italy\label{aff75}
\and
Universit\"at Bonn, Argelander-Institut f\"ur Astronomie, Auf dem H\"ugel 71, 53121 Bonn, Germany\label{aff76}
\and
INFN-Sezione di Roma, Piazzale Aldo Moro, 2 - c/o Dipartimento di Fisica, Edificio G. Marconi, 00185 Roma, Italy\label{aff77}
\and
Universit\'e C\^{o}te d'Azur, Observatoire de la C\^{o}te d'Azur, CNRS, Laboratoire Lagrange, Bd de l'Observatoire, CS 34229, 06304 Nice cedex 4, France\label{aff78}
\and
Universit\'e Paris Cit\'e, CNRS, Astroparticule et Cosmologie, 75013 Paris, France\label{aff79}
\and
Institut d'Astrophysique de Paris, 98bis Boulevard Arago, 75014, Paris, France\label{aff80}
\and
Institut de F\'{i}sica d'Altes Energies (IFAE), The Barcelona Institute of Science and Technology, Campus UAB, 08193 Bellaterra (Barcelona), Spain\label{aff81}
\and
School of Mathematics, Statistics and Physics, Newcastle University, Herschel Building, Newcastle-upon-Tyne, NE1 7RU, UK\label{aff82}
\and
Department of Physics and Astronomy, University of Aarhus, Ny Munkegade 120, DK-8000 Aarhus C, Denmark\label{aff83}
\and
Space Science Data Center, Italian Space Agency, via del Politecnico snc, 00133 Roma, Italy\label{aff84}
\and
Centre National d'Etudes Spatiales -- Centre spatial de Toulouse, 18 avenue Edouard Belin, 31401 Toulouse Cedex 9, France\label{aff85}
\and
Institute of Space Science, Str. Atomistilor, nr. 409 M\u{a}gurele, Ilfov, 077125, Romania\label{aff86}
\and
Departamento de Astrof\'isica, Universidad de La Laguna, 38206, La Laguna, Tenerife, Spain\label{aff87}
\and
Dipartimento di Fisica e Astronomia "G. Galilei", Universit\`a di Padova, Via Marzolo 8, 35131 Padova, Italy\label{aff88}
\and
Institut f\"ur Theoretische Physik, University of Heidelberg, Philosophenweg 16, 69120 Heidelberg, Germany\label{aff89}
\and
Institut de Recherche en Astrophysique et Plan\'etologie (IRAP), Universit\'e de Toulouse, CNRS, UPS, CNES, 14 Av. Edouard Belin, 31400 Toulouse, France\label{aff90}
\and
Universit\'e St Joseph; Faculty of Sciences, Beirut, Lebanon\label{aff91}
\and
Departamento de F\'isica, FCFM, Universidad de Chile, Blanco Encalada 2008, Santiago, Chile\label{aff92}
\and
Satlantis, University Science Park, Sede Bld 48940, Leioa-Bilbao, Spain\label{aff93}
\and
Infrared Processing and Analysis Center, California Institute of Technology, Pasadena, CA 91125, USA\label{aff94}
\and
Instituto de Astrof\'isica e Ci\^encias do Espa\c{c}o, Faculdade de Ci\^encias, Universidade de Lisboa, Tapada da Ajuda, 1349-018 Lisboa, Portugal\label{aff95}
\and
Universidad Polit\'ecnica de Cartagena, Departamento de Electr\'onica y Tecnolog\'ia de Computadoras,  Plaza del Hospital 1, 30202 Cartagena, Spain\label{aff96}
\and
INFN-Bologna, Via Irnerio 46, 40126 Bologna, Italy\label{aff97}
\and
Kapteyn Astronomical Institute, University of Groningen, PO Box 800, 9700 AV Groningen, The Netherlands\label{aff98}
\and
Dipartimento di Fisica, Universit\`a degli studi di Genova, and INFN-Sezione di Genova, via Dodecaneso 33, 16146, Genova, Italy\label{aff99}
\and
INAF, Istituto di Radioastronomia, Via Piero Gobetti 101, 40129 Bologna, Italy\label{aff100}
\and
Astronomical Observatory of the Autonomous Region of the Aosta Valley (OAVdA), Loc. Lignan 39, I-11020, Nus (Aosta Valley), Italy\label{aff101}
\and
ICL, Junia, Universit\'e Catholique de Lille, LITL, 59000 Lille, France\label{aff102}
\and
ICSC - Centro Nazionale di Ricerca in High Performance Computing, Big Data e Quantum Computing, Via Magnanelli 2, Bologna, Italy\label{aff103}
\and
Instituto de F\'isica Te\'orica UAM-CSIC, Campus de Cantoblanco, 28049 Madrid, Spain\label{aff104}
\and
CERCA/ISO, Department of Physics, Case Western Reserve University, 10900 Euclid Avenue, Cleveland, OH 44106, USA\label{aff105}
\and
Laboratoire Univers et Th\'eorie, Observatoire de Paris, Universit\'e PSL, Universit\'e Paris Cit\'e, CNRS, 92190 Meudon, France\label{aff106}
\and
Istituto Nazionale di Fisica Nucleare, Sezione di Ferrara, Via Giuseppe Saragat 1, 44122 Ferrara, Italy\label{aff107}
\and
Universit\'e de Strasbourg, CNRS, Observatoire astronomique de Strasbourg, UMR 7550, 67000 Strasbourg, France\label{aff108}
\and
Kavli Institute for the Physics and Mathematics of the Universe (WPI), University of Tokyo, Kashiwa, Chiba 277-8583, Japan\label{aff109}
\and
Dipartimento di Fisica - Sezione di Astronomia, Universit\`a di Trieste, Via Tiepolo 11, 34131 Trieste, Italy\label{aff110}
\and
NASA Ames Research Center, Moffett Field, CA 94035, USA\label{aff111}
\and
Bay Area Environmental Research Institute, Moffett Field, California 94035, USA\label{aff112}
\and
Institute Lorentz, Leiden University, Niels Bohrweg 2, 2333 CA Leiden, The Netherlands\label{aff113}
\and
Institute for Astronomy, University of Hawaii, 2680 Woodlawn Drive, Honolulu, HI 96822, USA\label{aff114}
\and
Department of Physics \& Astronomy, University of California Irvine, Irvine CA 92697, USA\label{aff115}
\and
Department of Astronomy \& Physics and Institute for Computational Astrophysics, Saint Mary's University, 923 Robie Street, Halifax, Nova Scotia, B3H 3C3, Canada\label{aff116}
\and
Departamento F\'isica Aplicada, Universidad Polit\'ecnica de Cartagena, Campus Muralla del Mar, 30202 Cartagena, Murcia, Spain\label{aff117}
\and
Department of Physics, Oxford University, Keble Road, Oxford OX1 3RH, UK\label{aff118}
\and
CEA Saclay, DFR/IRFU, Service d'Astrophysique, Bat. 709, 91191 Gif-sur-Yvette, France\label{aff119}
\and
Institute of Cosmology and Gravitation, University of Portsmouth, Portsmouth PO1 3FX, UK\label{aff120}
\and
Department of Computer Science, Aalto University, PO Box 15400, Espoo, FI-00 076, Finland\label{aff121}
\and
Ruhr University Bochum, Faculty of Physics and Astronomy, Astronomical Institute (AIRUB), German Centre for Cosmological Lensing (GCCL), 44780 Bochum, Germany\label{aff122}
\and
DARK, Niels Bohr Institute, University of Copenhagen, Jagtvej 155, 2200 Copenhagen, Denmark\label{aff123}
\and
Instituto de Astrof\'isica de Canarias (IAC); Departamento de Astrof\'isica, Universidad de La Laguna (ULL), 38200, La Laguna, Tenerife, Spain\label{aff124}
\and
Universit\'e PSL, Observatoire de Paris, Sorbonne Universit\'e, CNRS, LERMA, 75014, Paris, France\label{aff125}
\and
Universit\'e Paris-Cit\'e, 5 Rue Thomas Mann, 75013, Paris, France\label{aff126}
\and
Univ. Grenoble Alpes, CNRS, Grenoble INP, LPSC-IN2P3, 53, Avenue des Martyrs, 38000, Grenoble, France\label{aff127}
\and
Department of Physics and Astronomy, Vesilinnantie 5, University of Turku, 20014 Turku, Finland\label{aff128}
\and
Serco for European Space Agency (ESA), Camino bajo del Castillo, s/n, Urbanizacion Villafranca del Castillo, Villanueva de la Ca\~nada, 28692 Madrid, Spain\label{aff129}
\and
ARC Centre of Excellence for Dark Matter Particle Physics, Melbourne, Australia\label{aff130}
\and
Centre for Astrophysics \& Supercomputing, Swinburne University of Technology,  Hawthorn, Victoria 3122, Australia\label{aff131}
\and
School of Physics and Astronomy, Queen Mary University of London, Mile End Road, London E1 4NS, UK\label{aff132}
\and
Department of Physics and Astronomy, University of the Western Cape, Bellville, Cape Town, 7535, South Africa\label{aff133}
\and
ICTP South American Institute for Fundamental Research, Instituto de F\'{\i}sica Te\'orica, Universidade Estadual Paulista, S\~ao Paulo, Brazil\label{aff134}
\and
Oskar Klein Centre for Cosmoparticle Physics, Department of Physics, Stockholm University, Stockholm, SE-106 91, Sweden\label{aff135}
\and
Astrophysics Group, Blackett Laboratory, Imperial College London, London SW7 2AZ, UK\label{aff136}
\and
INAF-Osservatorio Astrofisico di Arcetri, Largo E. Fermi 5, 50125, Firenze, Italy\label{aff137}
\and
Dipartimento di Fisica, Sapienza Universit\`a di Roma, Piazzale Aldo Moro 2, 00185 Roma, Italy\label{aff138}
\and
Centro de Astrof\'{\i}sica da Universidade do Porto, Rua das Estrelas, 4150-762 Porto, Portugal\label{aff139}
\and
Institute of Astronomy, University of Cambridge, Madingley Road, Cambridge CB3 0HA, UK\label{aff140}
\and
Department of Astrophysics, University of Zurich, Winterthurerstrasse 190, 8057 Zurich, Switzerland\label{aff141}
\and
Theoretical astrophysics, Department of Physics and Astronomy, Uppsala University, Box 515, 751 20 Uppsala, Sweden\label{aff142}
\and
Department of Physics, Royal Holloway, University of London, Surrey TW20 0EX, UK\label{aff143}
\and
Department of Physics and Astronomy, University of California, Davis, CA 95616, USA\label{aff144}
\and
Department of Astrophysical Sciences, Peyton Hall, Princeton University, Princeton, NJ 08544, USA\label{aff145}
\and
Cosmic Dawn Center (DAWN)\label{aff146}
\and
Niels Bohr Institute, University of Copenhagen, Jagtvej 128, 2200 Copenhagen, Denmark\label{aff147}
\and
Center for Cosmology and Particle Physics, Department of Physics, New York University, New York, NY 10003, USA\label{aff148}
\and
Center for Computational Astrophysics, Flatiron Institute, 162 5th Avenue, 10010, New York, NY, USA\label{aff149}}    

\date{\today}

\abstract{
    We present \HS, a novel \texttt{Python} code to generate \Euclid realistic mock images in the \HE, \JE, \YE, and \IE\ photometric bands based on panchromatic {\it Hubble} Space Telescope observations. 
    The software was used to create a simulated database of \Euclid images for the 27 galaxy clusters observed during the Cluster Lensing And Supernova survey with {\it Hubble} (CLASH) and the Hubble Frontier Fields (HFF) program. Since the mock images were generated from real observations, they incorporate, by construction, all the complexity of the observed galaxy clusters. 
    The simulated \Euclid data of the galaxy cluster MACS~J0416.1$-$2403 were then used to explore the possibility of developing strong lensing models based on the \Euclid data. In this context, complementary photometric or spectroscopic follow-up campaigns are required to measure the redshifts of multiple images and cluster member galaxies. By Euclidising six parallel blank fields obtained during the HFF program, we provide an estimate of the number of galaxies detectable in \Euclid\ images per ${\rm deg}^2$ per magnitude bin (number counts) and the distribution of the galaxy sizes. Finally, we present a preview of the {\it Chandra} Deep Field South that will be observed during the Euclid Deep Survey and two examples of galaxy-scale strong lensing systems residing in regions of the sky covered by the Euclid Wide Survey. The methodology developed in this work lends itself to several additional applications, as simulated \Euclid\ fields based on HST (or JWST) imaging with extensive spectroscopic information can be used to validate the feasibility of legacy science cases or to train deep learning techniques in advance, thus preparing for a timely exploitation of the Euclid Survey data.
}
\titlerunning{\Euclid: Euclidised observations of Hubble Frontier Fields and CLASH galaxy clusters}
\authorrunning{Euclid Collaboration: P. Bergamini et
al.}
\maketitle
\section{Introduction}

Galaxy clusters, the most powerful gravitational lenses in the Universe, can distort dozens of background sources simultaneously, producing gravitational arcs and multiple images \citep[e.g.][]{Bergamini_2020}. Researchers routinely use these features to constrain the total mass distribution in cluster cores.

Various algorithms, including parametric and free form methods, have been developed for this task \citep[e.g.][]{Kneib_1993, Bradac_2004, Diego2005, lie06, Coe08, Jullo_lenstool, Zitrin_2009, Oguri_2010, Zitrin_2013, Lam2014}. Parametric models represent the cluster's total mass distribution as a collection of mass components, each characterised by parameters varied to fit the strong lensing constraints. These components account for cluster-scale dark matter haloes and the galaxy-scale substructure traced by cluster member galaxies. In contrast, the free-form approach often employs meshes or radial-basis functions to depict the cluster's total mass distributions. Several of these techniques are discussed and compared in~\cite{Meneghetti_2017}. \cite{Bergamini_2019} suggested combining strong lensing with stellar velocity dispersion measurements of cluster galaxies, obtained from the Multi Unit Spectroscopic Explorer \citep[MUSE, at the Very Large Telescope,][]{Bacon_MUSE} spectra, to enhance the precision of mass models on smaller scales~\citep[see also][]{Bergamini_2020, Meneghetti_2020, Granata_2021, Meneghetti_2022, Meneghetti_2023}. These additional data help reduce degeneracies between large- and small-scale cluster mass components. Accurate reconstruction of the inner cluster mass distribution is essential for enabling a variety of astrophysical and cosmological applications of strong lensing by galaxy clusters, including studying the nature of dark matter, examining the interplay between baryons and dark matter, exploring cosmic structure formation and evolution, constraining cosmological parameters, and utilising galaxy clusters as cosmic telescopes \citep[see][for extensive reviews]{Kneib_2011, Meneghetti_2013, Bartelmann_2013, Moresco_2022}.

Until recently, only a few tens of galaxy clusters could be accurately modelled for their total mass distributions. Strong lensing modelling requires high spatial resolution and deep observations, previously achievable mainly by the {\it Hubble} Space Telescope (HST) and, more recently, the {\it James Webb} Space Telescope (JWST). Most known strong lensing clusters were identified in ground-based observations \citep[e.g.][]{Bayliss_2011}, or by following up on galaxy clusters pre-selected based on their X-ray emission or the Sunyaev--Zeldovich effect amplitude \citep{Ebeling_2001, Ebeling_2007, Ebeling_2010, Ade_2016}. Confirming strong lensing features and measuring their redshifts required additional data from the HST (or JWST) and spectrographs on large telescopes.

This situation is set to change dramatically. Since last year, \Euclid has been observing most of the extragalactic sky with spatial resolution in the \IE\ band comparable to that of HST \citep{Laureijs11, Scaramella-EP1}. The Euclid Wide Survey (EWS) aims to cover about $15\,000\,\deg^2$ of the extragalactic sky to a minimum depth of $m_{\rm AB} = 24.5$ mag in the \IE band \citep{Cropper_2016}, with a signal-to-noise ratio (S/N) of 10 for extended sources. \Euclid will also survey the same area in the near-infrared \YE, \JE, and \HE\ bands \citep{Maciaszek22} to a depth of $m_{\rm AB} = 24.0$ mag with a minimum S/N of 5 for point sources. Additionally, its slitless spectroscopy will detect line emission with a sensitivity of $f_{{\rm H}\alpha} \gtrsim 2 \times 10^{-16}\,{\rm erg}\,{\rm s}^{-1} {\rm cm}^{-2}$ and a S/N of 3.5 for a typical source of size \ang{;;0.5}.

\begin{figure*}[t!]
\centering
\includegraphics[width=1\linewidth]{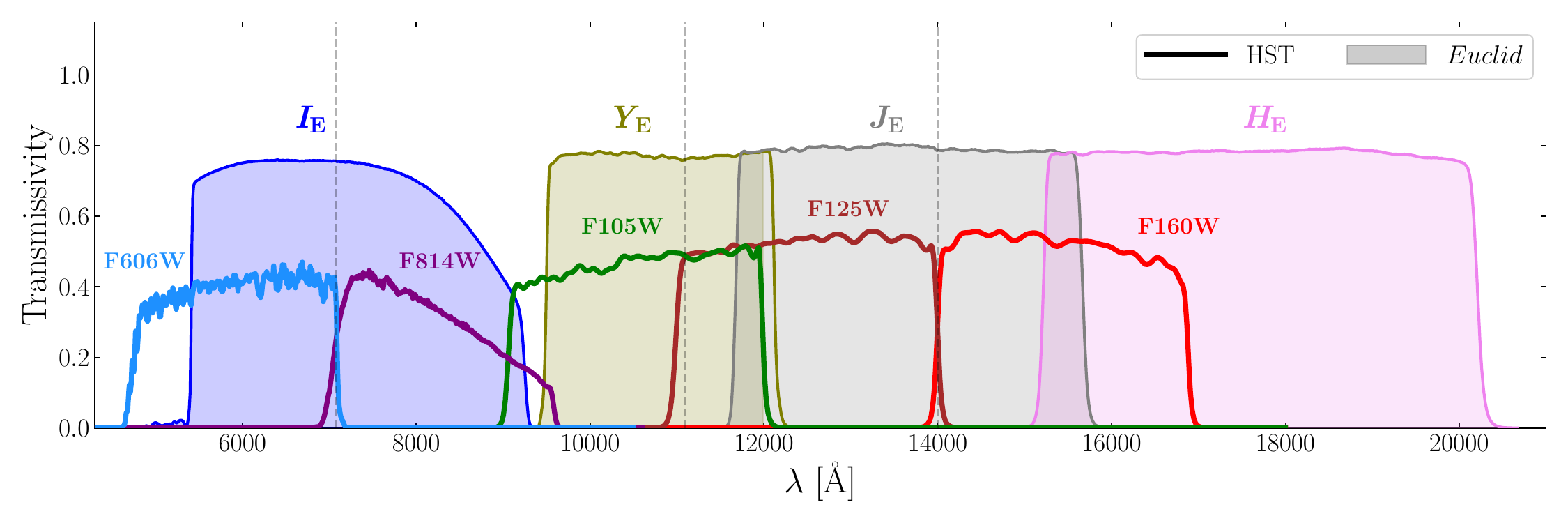}
\caption{Transmission curves of HST and \Euclid filters considered in this work. The five HST filters F606W, F814W, F105W, F125W, and F160W are shown with coloured solid lines, while the four \Euclid bands \IE, \YE, \JE, and \HE\ are represented by the coloured areas. The vertical black dashed lines indicate the wavelengths where the response of the HST F814W, F125W, and F160W filters begins to dominate over the F606W, F105W, and F125W filters, respectively.}
\label{fig:filters}
\end{figure*}

The primary goal of \Euclid is to reveal the nature of dark energy and dark matter, the two dominant components of the Universe, using weak lensing and galaxy clustering as the main probes. However, \Euclid will also significantly contribute to the discovery of galaxy clusters. During its lifetime, \Euclid is expected to observe over 60\,000 galaxy clusters \citep{Sartoris_2016}, with a S/N greater than 3, in the redshift range $0.2\le z \le 2$. About 5000 of these clusters are anticipated to be strong gravitational lenses containing multiple families of lensed images of background sources and strongly distorted radial and tangential arcs \citep{Boldrin_2012, Boldrin_2016}. Developing accurate strong lensing models for even a subset of this large number of cluster lenses will provide critical insights into the nature of dark matter and the growth of cosmic structures, provided that a large number of multiple images and cluster member galaxies can be identified from \Euclid observations.

This work introduces a tool to convert HST observations into \Euclid-like imaging data. We focus on HST observations of massive galaxy clusters collected in the Cluster Lensing and Supernova Survey with Hubble (CLASH,\footnote{\url{https://archive.stsci.edu/prepds/clash/}} \citealt{Postman_2012_clash}) and the Hubble Frontier Fields (HFF,\footnote{\url{https://archive.stsci.edu/prepds/frontier/}} \citealt{Lotz_2014HFF, Lotz_2017HFF}) programmes. We produced a simulated dataset from these images, consisting of mock observations of the same clusters in the EWS. We used simulated observations for the galaxy cluster MACS~J0416.1$-$2403 (hereafter, MACS~J0416) to quantify the number of strong lensing features detectable in future \Euclid observations of lens galaxy clusters. These are then used to preliminarily test the strengths and weaknesses of \Euclid-based cluster strong lensing models. Finally, the HFF parallel fields are employed to estimate the magnitude and size distributions of the luminous sources detectable in the \Euclid images.

The paper is organised as follows. In \Sec\ref{sec:Data}, we present the HST observational dataset used as input for the simulations. The simulation pipeline is detailed in \Sec\ref{sec:pipeline}. In \Sec\ref{sec:sim_val}, we describe various tests conducted on the mock \Euclid\ images to validate the simulations. In \Sec\ref{sec:lensing}, we present \Euclid-based strong lensing models for the galaxy cluster MACS~J0416, and the results of the strong lensing analysis are reported in \Sec\ref{sec:res}. In \Sec\ref{sec:applications}, we discuss other applications of the \Euclid\ simulated images, and the main conclusions of this work are outlined in \Sec\ref{sec:conclusions}. Throughout this work, magnitudes are given in the AB system \citep{Oke_1983}.

\begin{figure}[ht!]
\centering
\includegraphics[width=0.75\linewidth]{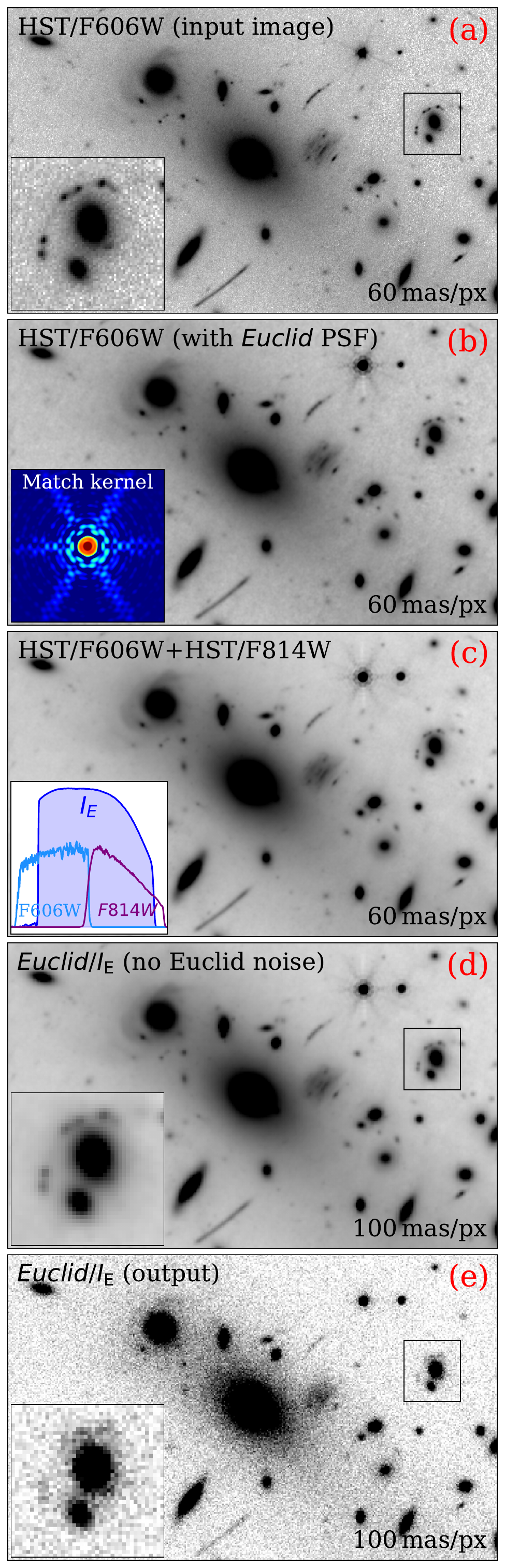}
\caption{Steps of the Euclidisation pipeline. In the top panel (a), we show the original {\it Hubble} F606W image of the galaxy cluster MACS~J0416 at $z=0.396$. The bottom panel (e) shows the final simulated \IE band. The intermediate panels (b), (c), and (d), from top to bottom, illustrate: the convolution to match the \Euclid PSF, the result of combining the {\it Hubble} F606W and F814W filters to produce the \IE band, and the re-binning necessary to match the \Euclid pixel scale in the \IE band (see \Sec\ref{sec:pipeline}).}
\label{fig:pipeline}
\end{figure}

\section{Input data}
\label{sec:Data}
The simulated \Euclid database of galaxy cluster observations presented in this work is based on the photometric data collected by the HST during the CLASH and HFF programmes. 
The former is a multi-cycle treasury programme that provided panchromatic images of 25 massive galaxy clusters in the redshift range [0.187, 0.890], for a total of 524 orbits from November 2010 to July 2013. The observations were carried out in 16 photometric bands ranging from UV to NIR wavelengths, using the Advanced Camera for Surveys (ACS, with filters: F435W, F475W, F606W, F625W, F775W, F814W, and F850W) and the Wide-Field Camera-3 (WFC3, with filters: F105W, F110W, F125W, F140W, F160W, F225W, F275W, F336W, and F390W).
The HFF programme provided deeper HST observations (840 HST orbits) of six massive clusters (four in common with CLASH), selected among the strongest known gravitational lenses, in seven ACS (F435W, F606W, and F814W) and WFC3 (F105W, F125W, F140W, and F160W) bands. 
The depth of the HFF images corresponds to a limiting magnitude of $m_{\rm AB} \sim 29$ with a S/N of 5 for point sources within an aperture of \ang{;;0.2} radius. This depth is about 1.5 magnitudes deeper than the CLASH observations. 

We note that the HFF and CLASH imaging is much deeper than the \Euclid\ observations we wish to produce, as discussed in the following sections. The images used in this work are drizzled to pixel scales of $65\,{\rm mas}\,{\rm px}^{-1}$ and $60\,{\rm mas}\,{\rm px}^{-1}$ for the CLASH and HFF data, respectively.  

\begin{figure*}[ht!]
\centering
	\includegraphics[width=0.9\linewidth]{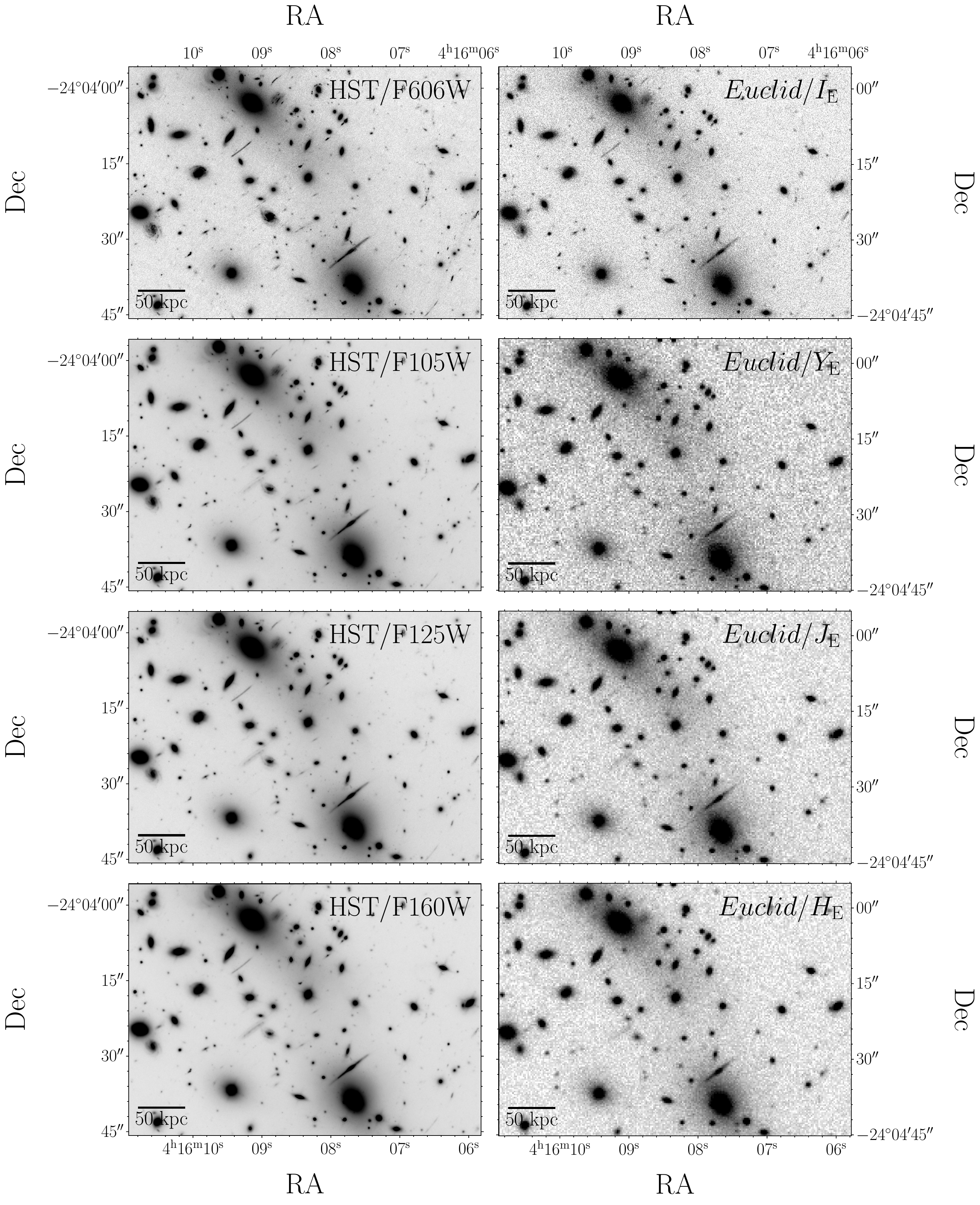}
	\caption{Images of the galaxy cluster MACS~J0416 in different photometric filters, observed with HST (left column) and \Euclid (right column). The \Euclid images are simulated using the pipeline described in \Sec\ref{sec:pipeline}, starting from the HST observations in the left column panels including the F814W filter, which is not displayed.}
	\label{fig:M0416_bands}
\end{figure*}

\begin{figure*}[ht!]
\centering
	\includegraphics[width=\linewidth]{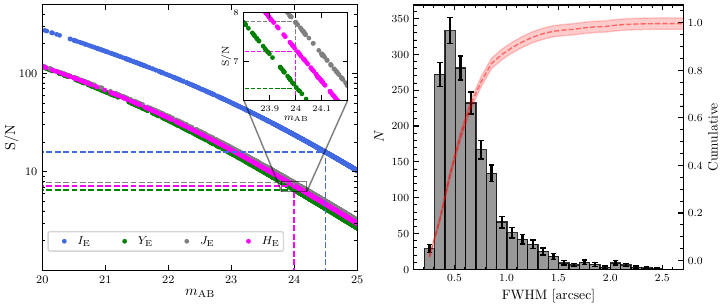}
	\caption{\emph{Left}: Signal-to-noise ratio (S/N) as a function of the magnitude for the luminous sources detected in the \Euclid\ simulated images of the galaxy cluster MACS~J0416. Blue, green, grey, and magenta dots indicate the S/N for galaxies detected in the \Euclid \IE, \YE, \JE, and \HE\ images, respectively, as a function of their magnitude in the different bands. These quantities are measured within circular apertures of radius $\ang{;;0.65}$ for the \IE\ band, and radius \ang{;;0.51} (equivalent in extension to the $\ang{;;0.9} \times \ang{;;0.9}$ aperture quoted in \T\,\ref{table:maglim_snrs}) for the \YE, \JE, and \HE\ filters. Vertical and horizontal dashed lines mark the values of the limiting S/Ns ($({\rm S/N})_{\rm lim}^{\IE}=15.9$, $({\rm S/N})_{\rm lim}^{\YE}=6.5$, $({\rm S/N})_{\rm lim}^{\JE}=7.8$, and $({\rm S/N})_{\rm lim}^{\HE}$=7.2) and limiting magnitudes ($m_{\mathrm{lim}}^{\IE}=24.5$, $m_{\mathrm{lim}}^{\YE}=24.0$, $m_{\mathrm{lim}}^{\JE}=24.0$, and $m_{\mathrm{lim}}^{\HE}=24.0$). \emph{Right}: Full width at half maximum (FWHM) distribution (grey) and its cumulative probability distribution (red) estimated for 1748 galaxies in six HFF parallel fields with \IE\ magnitude $\leq24.5$ (i.e. brighter than the limiting magnitude in the \IE\ filter).   
 }
	\label{fig:sn}
\end{figure*}

\section{Simulation pipeline}
\label{sec:pipeline}
This section describes the \texttt{Python}\footnote{\url{https://www.python.org/}} code \HS\ developed to simulate realistic \Euclid images from HST multi-band observations. Although our primary use in this work is the production of mock \Euclid observations of galaxy clusters, \HS\ is generally applicable to convert every sufficiently deep HST image into a \Euclid-like observation, with the only requirement being that the available HST bands overlap with the \Euclid photometric filters to be simulated (see \Fig\ref{fig:filters}). In particular, the HST F606W and F814W filters were used to simulate the \Euclid \IE\ band, while from the HST F105W, F125W, and F160W filters we simulate the \YE, \JE, and \HE\ bands (see \Sec\ref{sec:pipeline:combination}). In \Sec\ref{sec:res}, we present selected examples of \Euclid\ observations different from galaxy cluster fields. 

Figure\,\ref{fig:pipeline} visually summarises the different steps of the simulation pipeline (labelled a to e) implemented in the \HS\ code. These are described in detail in the following subsections, adopting the same letters as in \Fig\ref{fig:pipeline}, 
for clarity. 

\subsection{STEP a: Conversion of HST images to flux densities and image alignment}
\label{sec:pipeline:alligment}

In this first step of the Euclidisation pipeline,  we converted the pixel values of the HST images from units of electrons per second (${\rm e}\,{\rm s}^{-1}$) to physical flux densities (${\rm erg}\,{\rm s}^{-1}\,{\rm cm}^{-2}\,{\rm Hz}^{-1}$). The conversion was performed using the following equation:
\begin{equation}
f\mathrm{[{\rm erg}\,{\rm s}^{-1}\,{\rm cm}^{-2}\,{\rm Hz}^{-1}]}=F\mathrm{[{\rm e}\,{\rm s}^{-1}]} \times 10^{-0.4\,\left({{\rm ZP}_{\rm HST}+48.6}\right)}\, ,
    \label{eq:flux_density}
\end{equation}
where ${\rm ZP}_{\rm HST}$ is the instrumental HST zero point, defined as the AB magnitude of an object producing a signal of one electron per second in the HST image. The ${\rm ZP}_{\rm HST}$ value was computed from the pivot wavelength ({\tt PHOTPLAM}) and the inverse sensitivity ({\tt PHOTFLAM}), whose values are provided in the header of the HST Flexible Image Transport System (FITS, \citealt{pence2010}) files. Specifically, we used:\footnote{\url{www.stsci.edu/hst/instrumentation/acs/data-analysis/zeropoints}}
\begin{multline}
    {\rm ZP}_{\rm HST}=-2.5\logten\left(\texttt{PHOTFLAM}\right) \\
    - 5 \logten\left(\texttt{PHOTPLAM}\right) - 2.4079\, .
\end{multline}

Subsequently, we co-aligned the HST images in the different bands so that they shared the same pixel grid and astrometric calibration. For this purpose, we used the \texttt{Python} function \texttt{reproject\_exact} of the module \texttt{reproject}.\footnote{\url{https://reproject.readthedocs.io/en/stable/}}

\subsection{STEP b: Matching the HST and \Euclid point spread functions}
\label{sec:pipeline:psf}

The HST point spread functions (PSFs) have different shapes and full width at half maximum (FWHM) compared with the corresponding \Euclid\ PSFs.\footnote{The FWHM values of the HST PSFs in the F606W, F814W, F105W, F125W, and F160W filters range from approximately 70 mas to 130 mas. These values are smaller than those of the Euclid bands at comparable wavelengths (for example, \citealt{EuclidSkyOverview} report a measured PSF FWHM of 130 mas in the \IE\ band). Therefore, we can reliably match the HST resolution to that of Euclid by convolving the HST images with the appropriate matching kernels.} These differences must be taken into account to simulate realistic \Euclid\ observations. For each HST band, we derived a kernel function (hereafter, matching kernel) which, when convolved with the corresponding HST images, reproduced the correct PSF in the \Euclid\ simulated data.
 
The matching kernels between corresponding pairs of HST and \Euclid PSFs (see \Sec\,\ref{sec:pipeline:combination}) were computed using the \texttt{Python} function \texttt{create\_matching\_kernel} of the \texttt{Photutils} package \citep{photutils}.\footnote{\url{https://photutils.readthedocs.io/en/stable/psf_matching.html}} The HST PSF models were obtained with the \texttt{TinyTim} software.\footnote{\url{https://www.stsci.edu/hst/instrumentation/focus-and-pointing/focus/tiny-tim-hst-psf-modeling}} Although we verified that these PSFs yielded accurate results, \HS\ allows users full flexibility to implement alternative or customised PSF models. In panel (b) of \Fig\ref{fig:pipeline}, we show the effect of convolving the HST/F606W image of the galaxy cluster MACS~J0416 with the corresponding matching kernel (shown in the inset of the same panel) to the \IE\ band. 

\subsection{STEP c: Combination of HST filters to generate the \Euclid bands}
\label{sec:pipeline:combination}
In this step, we expressed the flux in a \Euclid\ band Q as a linear combination of $N$ fluxes in nearby HST filters $\{\mathrm{R}_j\}$:
\begin{equation}
\label{eq.: FQw}
F_\mathrm{Q} = \sum_{j=1}^N w_{j}\, F_{\mathrm{R}_j}=\vec{w}^\mathrm{T} \vec{F}_\mathbf{R}\, ,
\end{equation}
where the final term is expressed in linear algebra notation, with $\vec{w}$ and $\vec{F}_\mathbf{R}$ denoting column vectors. Below, we derive an expression for the weight vector $\vec{w}$. 

We assumed that the spectral density of flux $f(\lambda)$,--defined as the energy per unit time, per unit spherical angle, per unit detector area, per unit wavelength emitted by a source, (or $f(\nu)$ if it is expressed in terms of unit frequency)--could be described as a linear combination of $N$ base functions $\bigl\{ f_i(\lambda) \bigr\}$ (or $\bigl\{ f_i(\nu) \bigr\}$):
\begin{equation}
\label{eq.: flambda}
f(\lambda) = \sum_{i=1}^N k_i f_i(\lambda)=\frac{c}{\lambda^2}\sum_{i=1}^N k_i f_i(\nu)\, .
\end{equation}
Here, $k_i$ are constants associated with the decomposition of the spectral flux density. In computing the last term, we note that $f_i(\lambda)$ can be expressed as a function of frequency as
\begin{equation}
\label{eq.: fi}
    f_i(\lambda)=(c/\lambda^2)f_i(\nu)\, ,
\end{equation}
with $c$ being the light velocity in vacuum.
We can then express both $F_\mathrm{Q}$ and $F_{\mathrm{R}_j}$ as
\begin{equation}
\begin{aligned}
\label{eq.: FR_FQ}
&F_{\mathrm{R}_j} = \sum_i \mathcal{R}_{ij} k_i\; , \quad \mathrm{or} \quad \vec{F}_\mathbf{R} = \mathcal{R} \vec{k} \; ,
\\
&F_\mathrm{Q} = \sum_i q_i k_i=\vec{q}^\mathrm{T}\,\vec{k}\, ,
\end{aligned}
\end{equation}
where the matrix $\mathcal{R}$ is associated with the transmissivity of the HST filters, as detailed below. Assuming the AB magnitude system, all quantities above can be computed as (see \Eq 4 of \citealt{Hogg_2002}) 
\begin{align}
    \mathcal{R}_{ij} &= \frac{1}{c\, g^\mathrm{AB}_\nu}\frac{\int \mathrm{d}\lambda\, \lambda\, f_i(\lambda)\, R_j(\lambda)}{\int \frac{\mathrm{d}\lambda} {\lambda}\, R_j(\lambda)},\\
    q_i &= \frac{1}{c\, g^\mathrm{AB}_\nu}\frac{\int \mathrm{d}\lambda\, \lambda\, f_i(\lambda)\, Q(\lambda)}{\int \frac{\mathrm{d}\lambda} {\lambda}\, Q(\lambda)},
\end{align}
where $g^\mathrm{AB}_\nu$ is the zero point for the AB magnitude system ($g^\mathrm{AB}_\nu = 3631\, \mathrm{Jy}$, where $1\,\mathrm{Jy}=10^{-26}\,\mathrm{W}\,\mathrm{m}^{-2}\,\mathrm{Hz}^{-1}$), and $R_j(\lambda)$ and $Q(\lambda)$ are the transmissivities of the HST and \Euclid\ photometric bandpasses, respectively (see \Fig\ref{fig:filters}). 

By combining the two expressions in \Eq\eqref{eq.: FR_FQ}, we obtained\footnote{A necessary condition for the matrix $\mathcal{R}$ to be invertible is that the number of input filters has to be equal to the number of $f_i(\lambda)$ functions in \Eq\eqref{eq.: flambda} (i.e. $\mathcal{R}$ has to be a square matrix).
}
\begin{equation}
\label{Eq.: FQ}
F_\mathrm{Q} = \vec{q}^\mathrm{T}\,\vec{k}=\vec{q}^\mathrm{T}\,\mathcal{R}^{-1}\,\vec{F}_\mathbf{R} \, .
\end{equation}
Finally, by inserting this equation into \Eq\eqref{eq.: FQw}, we derived the following general expression for the weights:
\begin{equation}
\vec{w} = (\vec{q}^\mathrm{T}\, \mathcal{R}^{-1})^\mathrm{T} = \mathcal{R}^{-\mathrm{T}} \vec{q} \, .
\end{equation}

In \Fig\ref{fig:filters}, we present the transmission functions of the \Euclid\ photometric bands (\IE, \YE, \JE, and \HE, \citealt{Schirmer-EP18}) along with the HST filters used for simulations that cover a similar wavelength range to \Euclid\ (F606W, F814W, F105W, F125W, and F160W). The figure highlights that the wavelength range of the \YE\ band is well aligned with the HST F105W filter, while the \HE\ band overlaps solely with the F160W filter. For both cases, we assumed that the spectral density of flux per unit frequency, $f(\nu)$, of the sources in the images remains constant over the entire wavelength range covered by the input ($\mathrm{R}_1$) and output ($\mathrm{Q}_1$) filters, i.e. $f(\lambda)\propto c/\lambda^2$. For the \HE\ band, this assumption, necessitated by the absence of HST data covering the reddest part of the \HE\ band (see \Fig\ref{fig:filters}), is crucial to correctly recovering the measured magnitudes of the luminous sources in the simulated field. However, this represents a rough approximation of the true flux in the real \HE\ band, as it lacks photometric information from the source spectra at $\lambda>17\,000\,\AA$. Notably, Balmer break galaxies at redshift $z>3$ would appear brighter in \HE\ than in the F160W band. Under the previous assumption, $\mathcal{R}$ and $\vec{q}$ are one-dimensional matrices with $\mathcal{R}=\vec{q}=\mathrm{const.}$ and $\vec{w}=1$. Thus, the magnitudes of the sources in the \YE\ and \HE\ \Euclid\ bands are identical to those in the F105W and F160W HST filters, respectively.

In contrast, the \IE\ and \JE\ transmission functions span two HST bands. Specifically, the \IE\ band overlaps with both the F606W and F814W filters, while the \JE\ band overlaps with the F125W and F160W filters. In these cases, the weights, $\vec{w}$, were computed by assuming that the spectral density of flux, $f(\lambda)$, could be modelled as a sum of two top-hat functions $f_{\{1,2\}}(\nu)$ (see \Eq\ref{eq.: flambda}). The $f_1(\nu)$ base function is assumed to be constant within one of the two input HST bands and zero outside, while the opposite holds for $f_2(\nu)$. In this scenario, $\mathcal{R}$ is a two-dimensional matrix with $\mathcal{R}\propto \mathbb{I}$, where $\mathbb{I}$ represents the identity matrix, and the weights, $w_j$, assume the following values:
\begin{equation}
\begin{aligned}
       w_{\mathrm{F606W} \to \IE}&= 0.542 \, , \\
       w_{\mathrm{F814W} \to \IE}&= 0.458 \, , \\
       w_{\mathrm{F125W} \to \JE}&= 0.617 \, , \\
       w_{\mathrm{F160W} \to \JE}&= 0.383 \, .    
\end{aligned}
\end{equation}
We note that, according to the previous hypotheses, the weights corresponding to the same \Euclid band sum to one. 
In panel (c) of \Fig\ref{fig:pipeline}, we show the result of combining the two HST F606W and F814W filters into a single image, obtained using \Eq\eqref{eq.: FQw}.

\begin{table}  
	\tiny
	\def\arraystretch{1.6}
	\centering  
         \caption{Relevant quantities of the EWS.}  
        \label{table:maglim_snrs} 
	\begin{tabular}{|c|c|c|c|c|c|>{\centering\arraybackslash}m{9cm}|}
	   \hline
	   \multicolumn{6}{|c|}{\textbf{\normalsize Euclid wide survey (EWS)}}\\[2pt] 
	   \hline
	   {\it \textbf{Euclid}} \textbf{band} & \boldmath{$t_{\rm exp}$} & \boldmath{$m_{\rm lim}$} & \boldmath{${\rm SN}_{\rm lim}$} & {\textbf{Aperture size}} & \boldmath{$N_{\rm ap}^{\rm px}$}  \\[2pt] 
	   \hline
	   \boldmath{\IE} & 2280\,s & 24.5 & 15.9 & $r=\ang{;;0.65}$ & 132.7
	   \cr
	   \hline
	   \boldmath{\YE} & 448\,s & 24 & 6.5 & $\ang{;;0.9} \times \ang{;;0.9}$ & 9
	   \cr
	   \hline
	   \boldmath{\JE} & 448\,s & 24 & 7.8 & $\ang{;;0.9} \times \ang{;;0.9}$ & 9
	   \cr
	   \hline
	   \boldmath{\HE} & 448\,s & 24 & 7.2 & $\ang{;;0.9} \times \ang{;;0.9}$ & 9
	   \cr
	   \hline
	\end{tabular}
    \tablefoot{The values are reported by \citealt{Scaramella-EP1}.The total exposure time for each \Euclid band is denoted by $t_{\rm exp}$, and ${\rm SN}_{\rm lim}$ indicates the S/N of a source with a magnitude $m_{\rm lim}$, measured within an aperture with a total area (in pixels) equal to $N_{\rm ap}^{\rm px}$. A circular aperture of radius  \ang{;;0.65} is used for \IE\ images, while square apertures of $\ang{;;0.9} \times \ang{;;0.9}$ are used for \YE, \JE, and \HE\ images.}
	\smallskip
\end{table}

\subsection{STEP d: Projection onto the \Euclid pixel grid}
\label{sec:Step_d}
In this step, we re-binned the images from the HST pixel grid to the correct \Euclid pixel scale. To this aim, we employed the \texttt{Python} function \texttt{reproject\_exact} introduced above. The resulting images have pixel scales of $100\,{\rm mas}\,{\rm px}^{-1}$ in the \IE\ band and $300\,{\rm mas}\,{\rm px}^{-1}$ in \YE, \JE, and \HE\ filters.

The re-binned images, expressed in units of physical flux, were then converted into units of electrons per second by inverting \Eq\eqref{eq:flux_density}. For this conversion, we assumed a zero point of  ${\rm ZP}_\mathrm{Euclid}=23.9\,{\rm mag}$ for all bands. The resulting surface brightness per pixel in the \Euclid band $X$ is hereafter denoted as $F_{X,{\rm rebin}}$. Panel (d) of \Fig\ref{fig:pipeline} illustrates the result of the re-binning procedure for the \IE\ band.

\begin{figure}[ht!]
\centering
	\includegraphics[width=\linewidth]{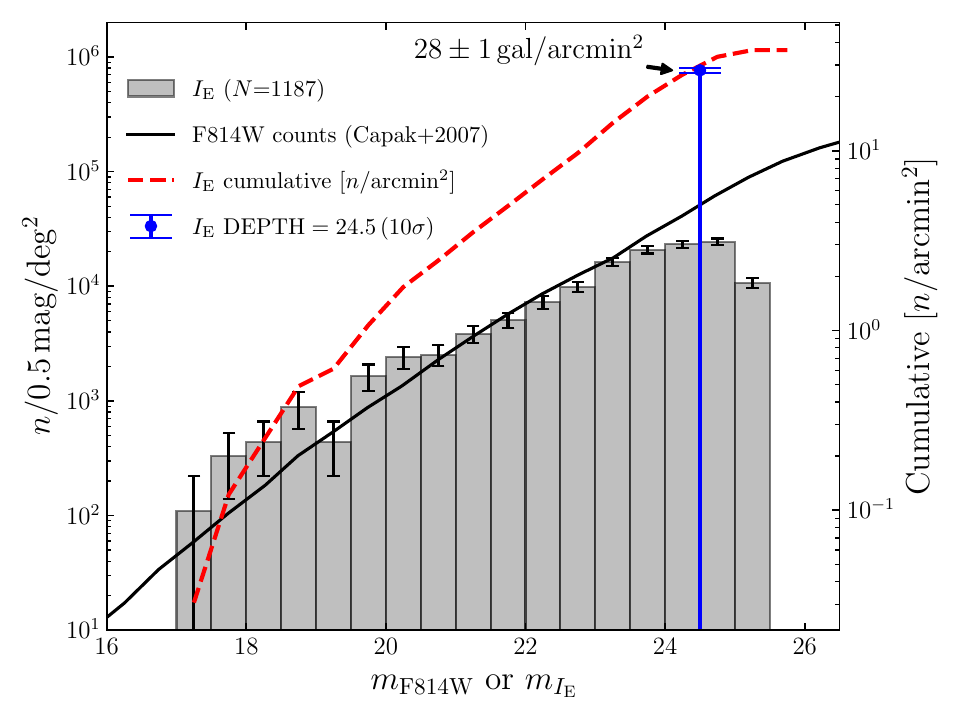}
	\caption{Galaxy number counts from \Euclid\ observations. The galaxy differential number counts derived from the simulated \Euclid observations in the \IE\ band of the six HFF parallel fields are shown by the grey histogram with error bars. For comparison, those measured from HST observations \citep[][]{Capak2007} are indicated by the black solid line. The vertical blue line marks the limiting magnitude in the \IE\ band ($24.5$ at $10\,\sigma$). The cumulative number of galaxies per ${\rm arcmin}^2$ is shown by the red dashed line. We estimate the number counts from a total area of $\sim32.67$\,arcmin$^2$. We quote the number of galaxies per ${\rm arcmin}^2$ at the limiting magnitude at the top of the blue vertical line.}
	\label{fig:numbercounts}
\end{figure}

\begin{figure*}[ht!]
\centering
	\includegraphics[width=1\linewidth]{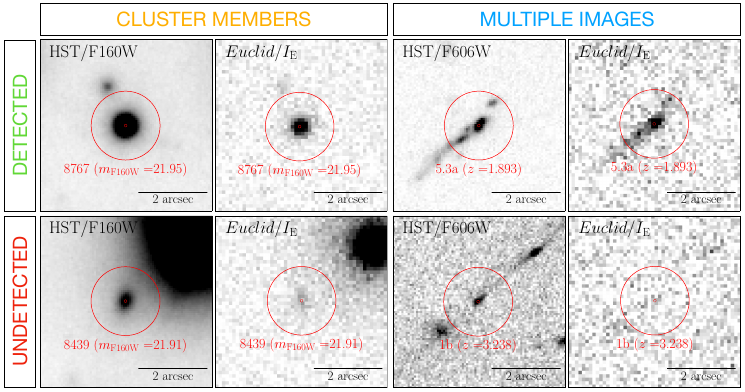}
	\caption{Comparison between cluster members and multiple images as observed by the HST and \Euclid. All sources are observed near the core of the galaxy cluster MACS~J0416. The upper panels show a cluster member and a lensed galaxy detected in both the HST F160W (HST F606W) and \Euclid \IE\ bands. In contrast, the sources in the bottom panels are detected in the HST data but not in the \Euclid images. In each panel, we report the IDs of the displayed cluster members and multiple images from the catalogues presented by \cite{Bergamini_M0416_2023}, the F160W total magnitude of the cluster galaxies, and the redshift of the lensed sources.}
	\label{fig:detections}
\end{figure*}

\subsection{STEP e: Noise addition}

The final step of the Euclidisation pipeline involved incorporating the appropriate noise into the \Euclid\ simulated images. To accomplish this, we assumed that the noise in the input HST background subtracted images was negligible compared to the final noise. This approximation is supported by the significantly greater depth of the HST observations compared to that of the \Euclid images.

The noise level was determined by adjusting the sky surface brightness of the \Euclid\ images to achieve the expected nominal S/N for a source with a specific limiting magnitude.
Specifically, to measure a S/N of ${\rm SN}_{\rm lim}$ for a source with flux of $F_{\rm lim}\,[{\rm e}\,{\rm s}^{-1}]$ measured within an aperture containing $N_{\rm ap}^{\rm px}$ pixels, the sky surface brightness per pixel in units of [${\rm e}\,{\rm s}^{-1}$] was calculated as
\begin{multline}
{\rm BK}_{\mathrm{sky}}[{\rm e}\,{\rm s}^{-1}] = \left\{\left(F_{\mathrm{lim}}[{\rm e}\,{\rm s}^{-1}]\right)^2 \times t_{\mathrm{exp}} \times \left({\rm SN}_{\mathrm{lim}}\right)^{-2}-F_{\mathrm{lim}}[{\rm e}\,{\rm s}^{-1}]\right\}\\
\times \left(N^{\mathrm{px}}_{\mathrm{ap}}\right)^{-1}\, ,
\end{multline}
where $t_{\mathrm{exp}}$ is the exposure time of the \Euclid\ observations. In the EWS, this corresponds to 2280\,s for the \IE\ band and 448\,s for the \YE, \JE, and \HE\ bands \citep{Scaramella-EP1}. The limiting flux, $F_{\mathrm{lim}}$, was computed from the limiting magnitude, $m_{\rm lim}$, using the zero point introduced in \Sec\ref{sec:Step_d}:
\begin{equation}
F_{\mathrm{lim}}[{\rm e}\,{\rm s}^{-1}] = 10^{0.4\left(\mathrm{ZP}_\mathrm{Euclid}-m_{\mathrm{lim}}\right)}\, .
\end{equation}

The expected limiting magnitudes and S/N in the EWS are reported in \T\ref{table:maglim_snrs} \citep{Scaramella-EP1}.
The noise was generated using a Poisson process, where the noise variance is equal to the sum of $F_{X,{\rm rebin}}$ and ${\rm BK}_{\mathrm{sky}}$.
Panel (e) of \Fig\ref{fig:pipeline} shows the resulting simulated \Euclid image in the \IE\ band, including the noise. 

\Fig\ref{fig:M0416_bands} presents the results of applying the simulation pipeline, from step (a) to (e), to create mock \Euclid observations in all the photometric bands for a region with an approximate size of $\ang{;;70} \times \ang{;;50}$ encompassing the core of the galaxy cluster MACS~J0416.
The left-hand panels show the input HST photometric data (excluding the HST F814W filter) while the right-hand panels show the output simulated \Euclid\ images in the \IE, \YE, \JE, and \HE\ bands. Despite the shallower depth of the EWS observations compared to the CLASH and HFF observations, many strong lensing features, including some prominent gravitational arcs, are clearly visible in the \Euclid\ images. Notably, the higher spatial resolution of the \IE\ band allowed us to distinguish small details, such as stellar clumps and spiral structures, which appear in some of the multiple images of background lensed galaxies. These features are fundamental in constraining the total mass distribution of galaxy clusters through strong gravitational lensing \citep[e.g.][]{Bergamini_2020,Pignataro_2021}, as discussed in subsequent sections. Simulated \Euclid\ images similar to those presented in this section were created for all 27 galaxy clusters observed during the CLASH and HFF programmes.  

\section{Simulation validation}
\label{sec:sim_val}
This section discusses three tests performed on the mock \Euclid\ images to validate the simulation pipeline. In the first test, we verified that the expected depth of the EWS observations was correctly reproduced in the simulations. In the second, we derived the distribution of the sizes of the galaxies in blank fields. Finally, in the third test, we estimated the number counts of galaxies detected in the \Euclid\ images. For tests two and three, we made use of simulated \IE\ images of six cluster-parallel blank fields obtained during the HFF observations.

The coloured dots in the left panel of \Fig\ref{fig:sn} show the measured S/N for the luminous sources identified in the different \Euclid\ bands of the galaxy cluster MACS~J0416, plotted as a function of their magnitude. These sources include cluster members and galaxies in the cluster foreground or background. Both the S/N and the magnitudes were measured within circular apertures containing 132.7 pixels for the \IE\ band and 9 pixels for the \YE, \JE, and \HE\ filters (see \T\ref{table:maglim_snrs}). 
The figure shows that for source magnitude values equal to the limiting magnitude, the measured S/N precisely matches the expected EWS value (dashed coloured lines; see also \T\ref{table:maglim_snrs}). This test demonstrates that the noise in the simulated \Euclid\ images reproduces the expected depth of the EWS data.

The right panel of \Fig\ref{fig:sn} shows the distribution of the circularised FWHM for 1748 galaxies, with \IE\ magnitudes $\leq24.5$ (i.e. down to the limiting magnitude), identified in the \IE\ images of the six Euclidised parallel fields described above. The cumulative probability distribution is also plotted in red. 
This analysis reveals that $93.8\%$ of the selected galaxies have ${\rm FWHM}\leq\ang{;;1.3}$, with a median value of \ang{;;0.59}. Assuming a Gaussian profile for the galaxy surface brightness distribution and a median FWHM equal to $\ang{;;0.59}$, we find that $96.4\%$ of the brightness of the galaxies is enclosed within an aperture of \ang{;;0.65} radius. This value corresponds to the size of the aperture used to measure the limiting magnitude and S/N in the \IE\ band (see \T\ref{table:maglim_snrs}).   

Finally, \Fig\ref{fig:numbercounts} shows the measured distribution of number counts, i.e. the number of detected galaxies per ${\rm deg}^2$ per magnitude bin, in the six HFF parallel fields. In the figure, the \Euclid\ number counts (grey histogram) are compared with those estimated from HST observations \citep[black curve,][]{Capak2007}. From the cumulative distribution of galaxies (red dashed line), we estimate that the density of resolved galaxies brighter than the limiting magnitude ($\IE$ magnitude $\le24.5$) is equal to $28\pm1$ galaxies per ${\rm arcmin}^2$, in full agreement with expectations for the EWS observations \citep{Laureijs11}.

\begin{figure}[t!]
\centering
	\includegraphics[width=1\linewidth]{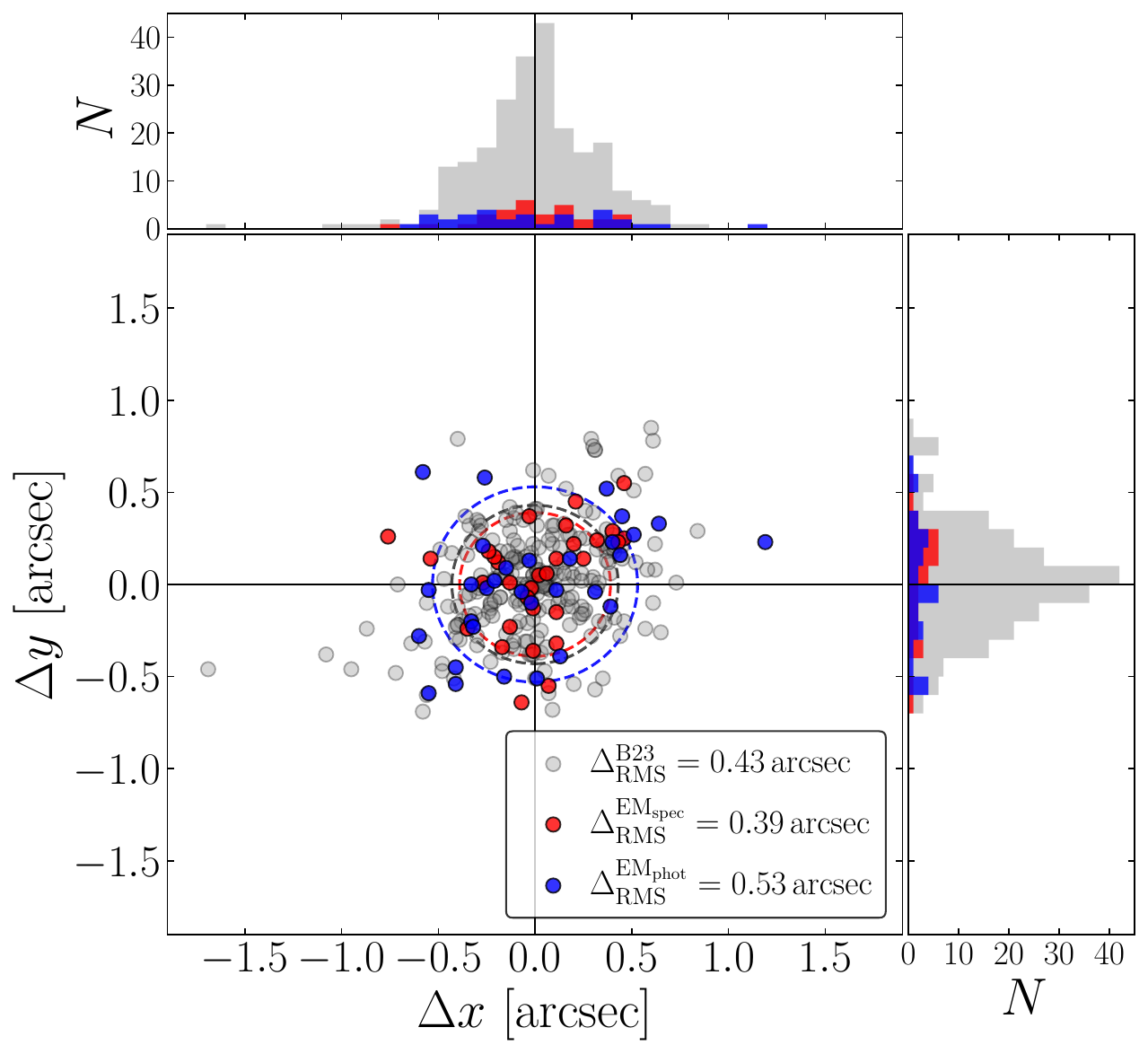}
	\caption{Displacements in the sky plane between the observed and model-predicted positions of the multiple images. We show the results for the models $\mathrm{EM_{spec}}$ and $\mathrm{EM_{phot}}$ in red and blue, respectively. For comparison, we also show the results obtained by B23 in grey. The histograms show the distribution of the displacements along the two directions. While the B23 model is based on 237 multiple images from 88 background sources, the $\mathrm{EM_{spec}}$ and $\mathrm{EM_{phot}}$ models are constructed using constraints from 31 multiple images of 12 sources. We quantify the accuracy of each model in terms of the $\Delta_{\rm RMS}$ (see \Sec\ref{sec:res}), as reported in the legend. This accuracy also determines the radii of the coloured dashed circles in the figure.}
	\label{fig:rms}
\end{figure}

\begin{figure*}[ht!]
\centering
	\includegraphics[width=1\linewidth]{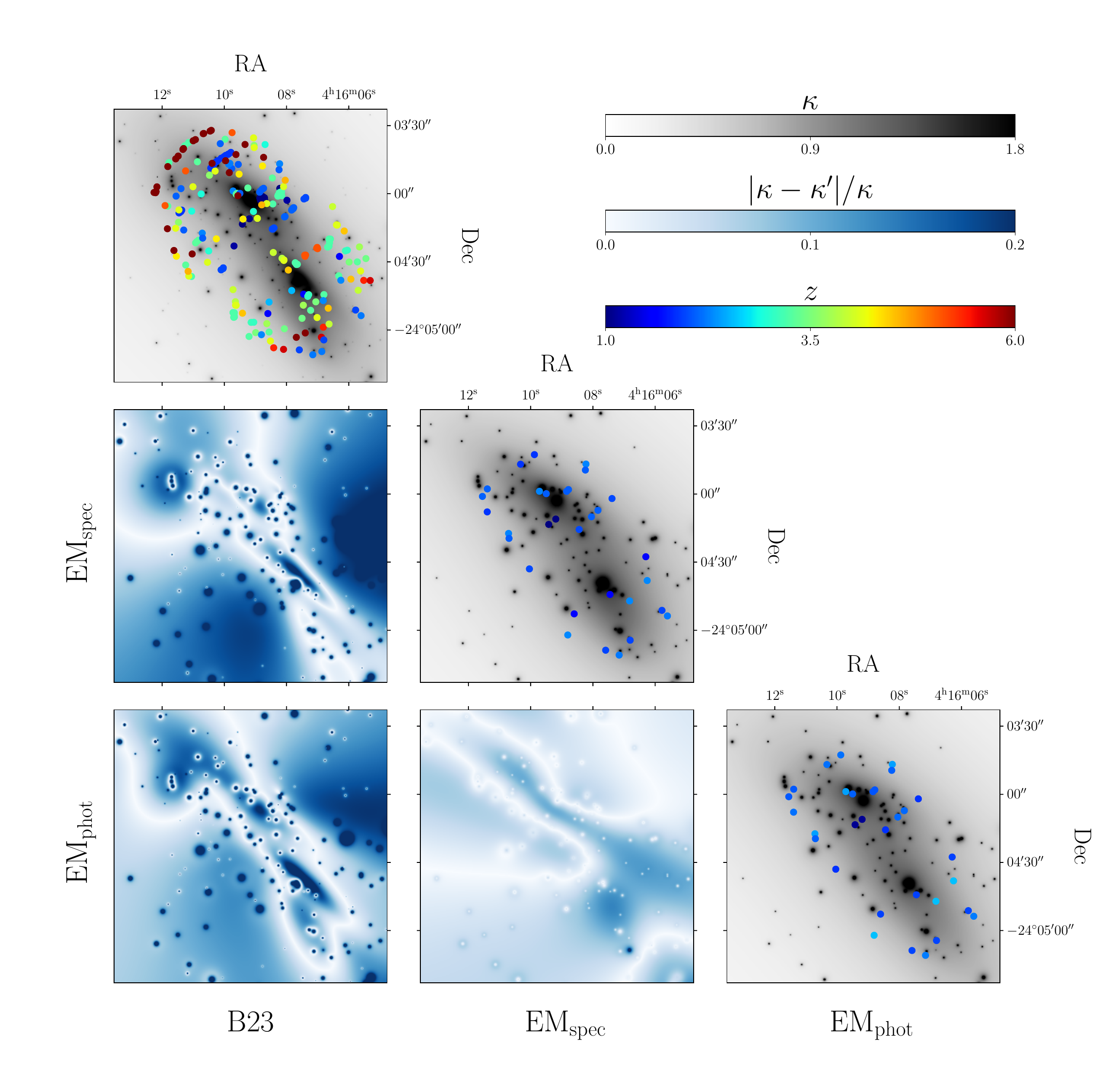}
	\caption{Comparison between the convergence maps for the galaxy cluster MACS~J0416, obtained from the B23, $\mathrm{EM_{spec}}$, and $\mathrm{EM_{phot}}$ strong lensing models. The maps are rescaled such that the ratio of the lens-source and observer-source angular diameter distances equals one. At the top of each column, we show the convergence maps obtained from the different models, while the dots mark the positions of the multiple images. The latter are colour-coded according to their redshifts. The panels at the intersections between pairs of models show the relative difference between the corresponding convergence maps.}
	\label{fig:kappa}
\end{figure*}

\begin{figure*}[ht!]
\centering
	\includegraphics[width=1\linewidth]{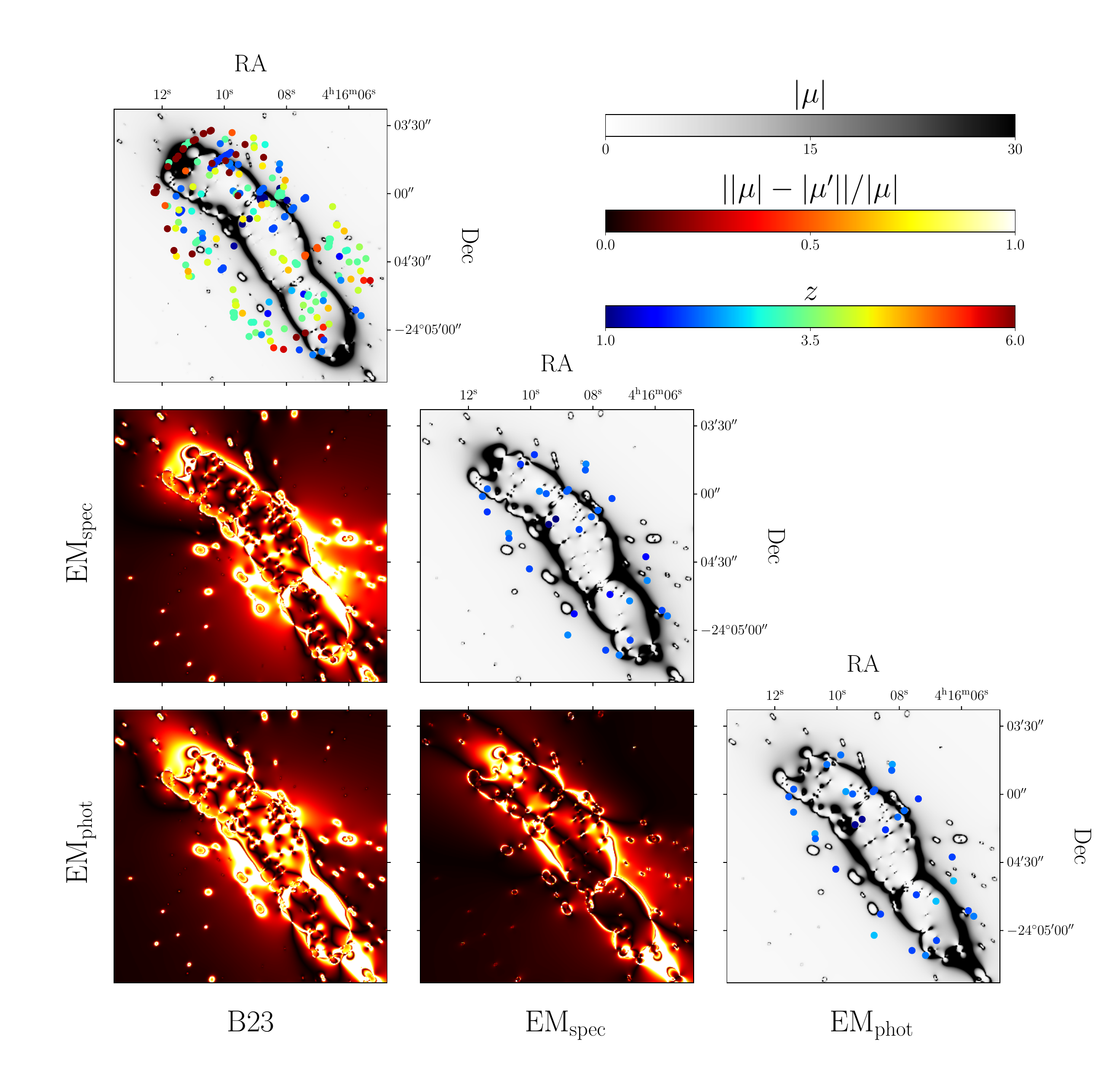}
	\caption{Comparison between the magnification maps of the galaxy cluster MACS~J0416, obtained from the B23, $\mathrm{EM_{spec}}$, and $\mathrm{EM_{phot}}$ strong lensing models.  We show the results for a source at redshift $z=3$. The model magnification maps are shown at the top of each column. The dots mark the positions of the multiple images, colour-coded according to their redshifts. The panels at the intersections between pairs of models show the relative differences between the corresponding magnification maps.}
	\label{fig:magnification}
\end{figure*}

\section{Parametric strong lensing models of galaxy clusters based on \Euclid observations}
\label{sec:lensing}

As a possible application of the simulated \Euclid photometric images, we present a preliminary analysis to quantify the precision and accuracy achievable by parametric strong lensing models of galaxy clusters based on \Euclid\ data. Our analysis focuses on the HFF galaxy cluster MACS~J0416, for which~\cite{Bergamini_M0416_2023} developed a high-precision strong lensing model (hereafter the B23 model), based on the spectrophotometric data obtained with HST and MUSE. Both the B23 model and the other lens models presented in this work are constructed using the publicly available parametric software \texttt{LensTool}~\citep{Jullo_lenstool, Limousin_lenstool, Jullo_Kneib_lenstool}.  

In the B23 model, the total projected gravitational potential of MACS~J0416 is expressed as the sum of several contributions, each corresponding to a different mass component: 
\begin{equation}
    \label{eq.: pot_dec}
    \phi_{\mathrm{tot}}= \sum_{i=1}^{N_{\rm h}}\phi_i^{\mathrm{halo}}+\sum_{j=1}^{N_{\rm G}}\phi_j^{\mathrm{gas}}+\sum_{k=1}^{N_{\rm g}}\phi_k^{\mathrm{gal}}+\phi^{\mathrm{fg}}\, .
\end{equation}
The $\phi_i^{\mathrm{halo}}$ terms represent the contribution from $N_{\rm h}=4$ cluster-scale haloes to the total cluster gravitational potential. Three of these haloes are parametrised as elliptical dual pseudo-isothermal mass distributions~\citep[dPIEs,][]{Limousin_lenstool, Eliasdottir_lenstool, Bergamini_2019} with an infinite truncation radius: two are centred close to the positions of the two brightest cluster galaxies (BCGs), and one is located in the southern part of the cluster, providing second-order corrections to the total mass distribution in that region. The fourth halo is a circular, non-truncated, dPIE profile that accounts for an overdensity of galaxies in the north-eastern region of the cluster.
In addition to these cluster-scale haloes, $N_{\rm G}=4$ dPIEs ($\phi_j^{\mathrm{gas}}$ in \Eq\ref{eq.: pot_dec}) are used to describe the hot gas content of the cluster. The values of the parameters of these profiles are fixed in the B23 lens model, as they were determined from the analysis of Chandra X-ray data of the cluster performed by~\cite{Bonamigo_2018}.
The sub-halo mass component of MACS~J0416 (corresponding to the terms $\phi_k^{\mathrm{gal}}$ in \Eq\ref{eq.: pot_dec}) comprises a total of $N_{\rm g}=213$ cluster member galaxies, including the two BCGs. Of these galaxies, 212 are parametrised as circular, core-less dPIEs,  whose central velocity dispersions ($\sigma_0$) and truncation radii ($r_{\mathrm{cut}}$) are scaled with their luminosities, following the two scaling relations reported in \Eq(4) of \cite{Bergamini_M0416_2023}. The remaining galaxy, identified as Gal-8971, is separately parametrised as an elliptical, core-less dPIE since its total mass is responsible for the formation of a galaxy-galaxy strong lensing system that creates four multiple images of a background source at $z=3.221$. Finally, the last term in \Eq\eqref{eq.: pot_dec}, $\phi^{\mathrm{fg}}$, accounts for the contribution to the lensing observables from a single foreground galaxy in the southwestern region of the cluster. This galaxy is described as a circular core-less dPIE. 

The optimal values of the free parameters of the profiles defined in \Eq\eqref{eq.: pot_dec} were determined by minimising the following $\chi^2$ function, which quantifies how well the lens model predicts the observed positions of the multiple images:
\begin{equation}
    \label{eq.: chi_lt}
    \chi^2(\pmb{\xi}) := \sum_{j=1}^{N_\mathrm{sou}} \sum_{i=1}^{N_{\mathrm{im}}^j} \left(\frac{\left\| \mathbf{x}_{i,j}^{\mathrm{obs}} - \mathbf{x}_{i,j}^{\mathrm{pred}} (\pmb{\xi}) \right\|}{\Delta x_{i,j}}\right)^2\, ,
\end{equation}
where $\Delta x_{i,j}$ are the uncertainties on the observed positions of the images, $N_{\mathrm{im}}^j$ is the number of multiple images of the same $j\mathrm{-th}$ background source, and $N_\mathrm{sou}$ is the total number of sources. The B23 lens model is constrained by the observed positions of 237 spectroscopically confirmed multiple images from 88 background sources within the redshift range $0.94 \leq z \leq 6.63$. 
We refer to \cite{Bergamini_M0416_2023} for a detailed description of the lens model.  

By exploiting the \Euclid\ simulated images of the galaxy cluster MACS~J0416 in the \IE, \YE, \JE, and \HE\ bands, we developed two \Euclid-based lens models, hereafter identified as $\mathrm{EM_{spec}}$ and $\mathrm{EM_{phot}}$. Both $\mathrm{EM_{spec}}$ and $\mathrm{EM_{phot}}$ assume a parametrisation for the total mass of MACS~J0416 similar to that adopted in the B23 model, but with the following three important differences. First,  the cluster-scale component of MACS~J0416 is parametrised by using just two non-truncated dPIE profiles centred on the positions of the BCGs. Second, the hot-gas mass component is not considered (i.e. the $\phi_j^{\mathrm{gas}}$ terms are not present in the models). Third, the sub-halo mass component of the cluster contains only those 125 cluster galaxies that are identifiable in the \Euclid \IE\ band and with an \IE\ magnitude $\leq22.5$ (see \Fig\ref{fig:detections}). All these galaxies are parametrised as circular, core-less dPIEs, adopting the $\sigma_0$--$L$ and $r_{\rm cut}$--$L$ scaling relations mentioned above, where the luminosity $L$ corresponds to the Kron magnitude of the galaxies measured in the \IE\ band (instead of the HST F160W Kron magnitude adopted in the B23 model). Since Gal-8971 now follows the scaling relations, the number of free parameters in the lens model is reduced by four. Thus, the \Euclid-based lens models count a total of 16 free parameters (12 associated with the parametric profiles used to describe the cluster-scale total mass distribution, two are the normalisations of the $\sigma_0$--$L$ and $r_{\rm cut}$--$L$ cluster member scaling relations, and two are used to parameterise the foreground galaxy residing in the south-western region of the cluster). We also note that, contrary to the B23 lens model, we did not assume any Gaussian prior on the normalisation of the $\sigma_0$--$L$ scaling relation. In the B23 model, this Gaussian prior is inferred from the measure stellar kinematics of the cluster member galaxies, through the procedure described by \cite{Bergamini_2019}.

The $\mathrm{EM_{spec}}$ and $\mathrm{EM_{phot}}$ were constrained by the observed positions of 31 multiple images from 12 background sources within the spectroscopic redshift interval $1.01 \leq z \leq 2.30$. This corresponds to the subsample of multiple images of the original B23 catalogue that were identifiable through visual inspecting the \Euclid\ \IE\ simulated observation. The larger pixel scale and lower resolution and depth of the \Euclid\ images compared to the original HST data allowed the secure detection of 13\% of the multiple images used in the B23 model. Figure \ref{fig:detections} shows examples of detected and non-detected cluster galaxies and multiple images in the \IE\ image.

The $\mathrm{EM_{spec}}$ and $\mathrm{EM_{phot}}$ models differ in that, while in the former we used the spectroscopic redshifts of the observed multiple images, in the latter we assumed photometric redshift measurements. To simulate the photometric redshift measurements, we randomly extracted, for each source, a redshift value from a Gaussian distribution centred on the source spectroscopic redshift, $z_{\rm spec}$, and with a standard deviation equal to $(1 + z_{\rm spec})\,0.05$. This corresponds to the expected uncertainty on photometric redshift measurements in the EWS \citep{Laureijs11,Desprez-EP10,EP-Paltani}.

We note that $\mathrm{EM_{spec}}$ and $\mathrm{EM_{phot}}$ are optimistic examples of lens models based on \Euclid data. To construct these models, we used the multiple images and cluster member catalogues from \citealt{Bergamini_M0416_2023}, based on HST photometric data and VIMOS and MUSE spectroscopic data, to identify the sources detectable in the \Euclid\ simulated observations. In more realistic cases, these components for the strong lensing models will be derived solely from the \Euclid data. For example, we will identify the cluster galaxies from the observations in the four \Euclid bands. \cite{Angora_2020} show that this task can be accomplished using convolutional neural networks (CNNs). The proposed technique will be tested with simulated \Euclid images obtained with \HS\ in an upcoming paper (Angora et al. in preparation). Spectroscopic follow-up observations will also be critical for identifying pure and complete samples of cluster members and candidate multiple images, as well as for measuring their redshifts. Despite these considerations, our results are informative regarding the precision and accuracy potentially achievable in strong lensing models based on \Euclid\ data. Previous studies, such as those by \cite{Johnson_2016} and \cite{Meneghetti_2017}, discuss how the accuracy and precision of strong lensing models depend on the availability of multiple images and spectroscopic redshifts.

\begin{figure}[t!]
\centering
	\includegraphics[width=1\linewidth]{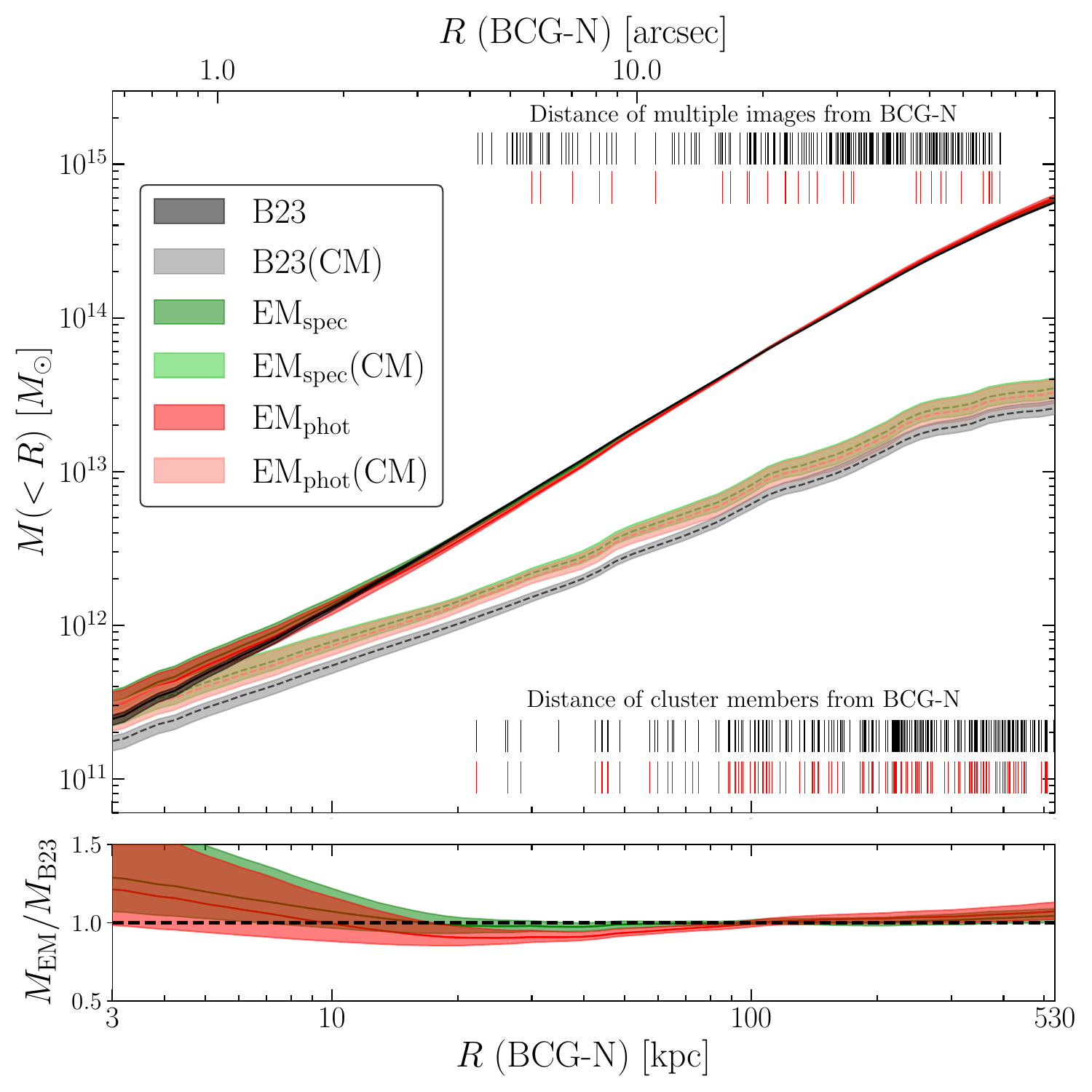}
	\caption{\emph{Top panel}: Cumulative total mass profiles of MACS~J0416 as derived from the B23, $\mathrm{EM_{spec}}$, and $\mathrm{EM_{phot}}$ strong lensing models. We show the results for the three models using dark grey, green, and red colours, respectively. The coloured bands indicate the 68.3\% confidence intervals. We also show the cumulative total mass profiles for the cluster member component (CM) using lighter colours. The radial distances are measured with respect to the BCG-N. We indicate the positions of multiple images and cluster members used in the \Euclid-based and B23 models with red and black vertical sticks at the top and bottom of the figure, respectively. \emph{Bottom panel}: Ratios between the cumulative total mass profiles derived from the \Euclid-based and the B23 models. The coloured bands indicate the 68.3\% confidence intervals.}
	\label{fig:mass_profile}
\end{figure}

\begin{figure*}[ht]
\centering
	\includegraphics[width=1\linewidth]{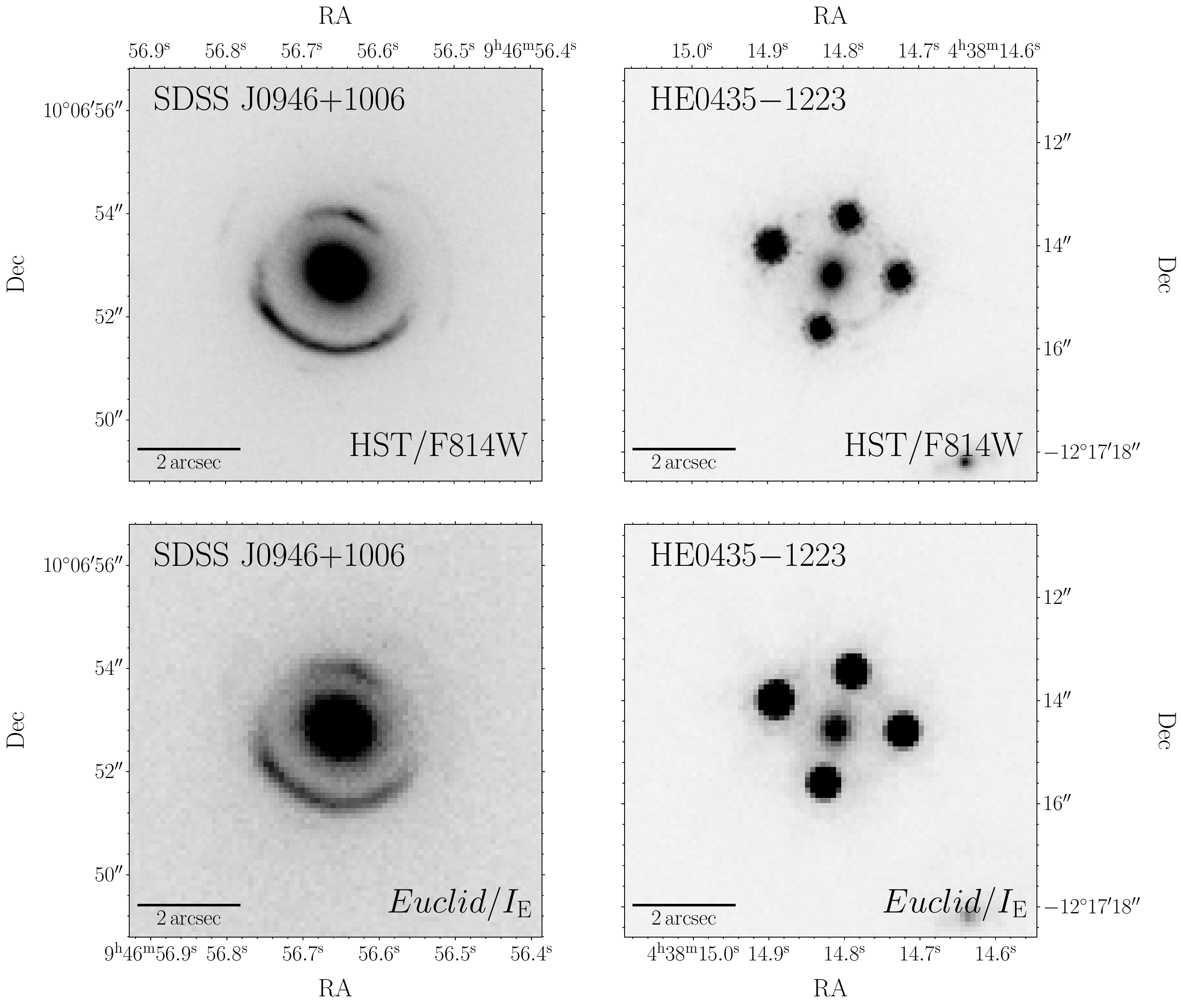}
	\caption{Examples of simulated \Euclid\ images of galaxy-galaxy strong lensing systems as observed during the \Euclid\ wide survey. In the upper panels, we show the two HST images used to generate the simulations in the lower panels. The system on the left, identified as SDSS~J0946$+$1006 (also known as the Jackpot lens), consists of a double Einstein ring lensed by a galaxy at $z=0.222$ (at the centre of the image). The HE0435$-$1223 system on the right is a quadruple imaged quasi-stellar object (QSO) at $z=1.693$ lensed by a galaxy at $z=0.455$.}
	\label{fig:ggsl}
\end{figure*}

\begin{figure*}[ht]
\centering
	\includegraphics[width=1\linewidth]{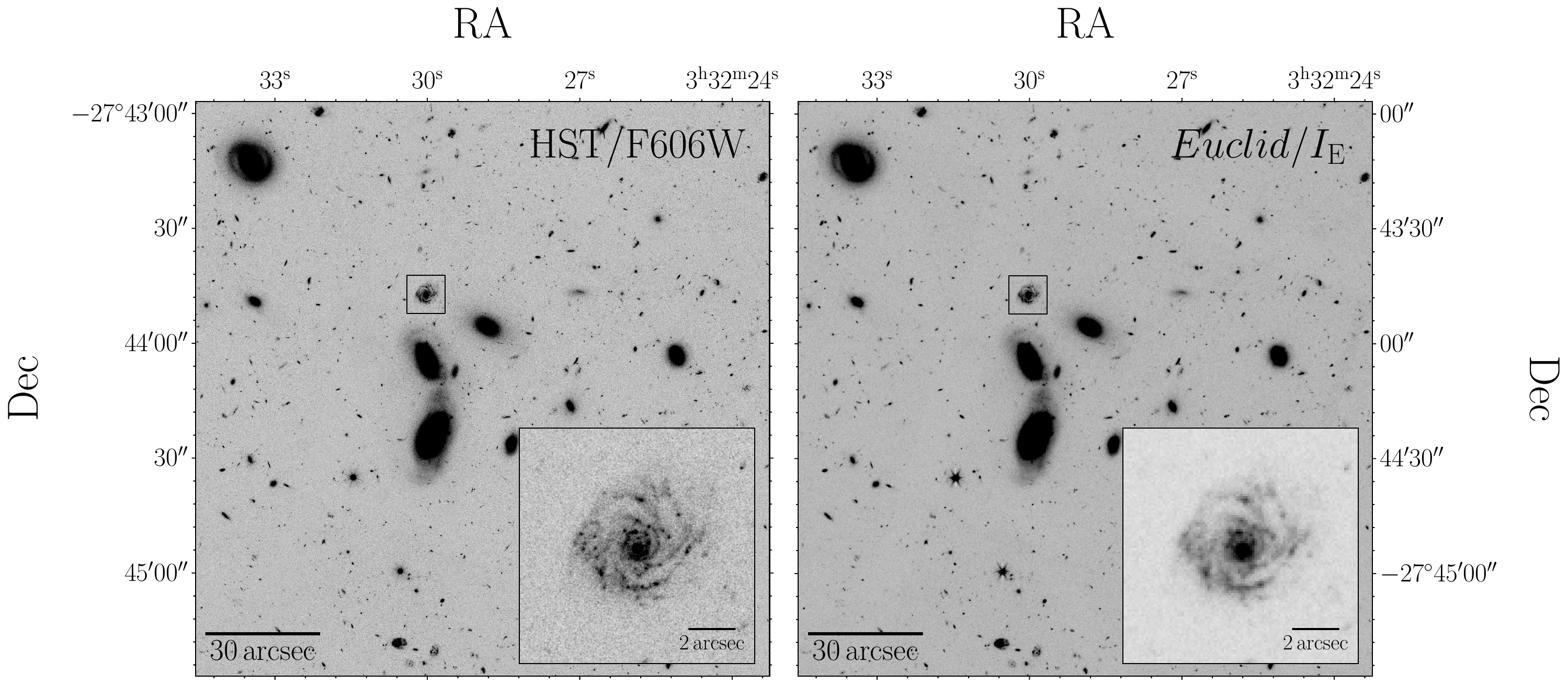}
	\caption{Comparison between a $\ang{;;150}\times\ang{;;150}$  portion of the {\it Chandra} Deep Field South as seen by HST (real data) or \Euclid (simulated). An exposure time of 90\,744\,s and a limiting magnitude of 26.5, corresponding to the expected depth of the \Euclid deep fields, is assumed in the simulation.}
	\label{fig:cdfs}
\end{figure*}

\section{Results of the strong lensing analysis on the simulated Euclid clusters}
\label{sec:res}

To accurately constrain the total mass distribution of galaxy clusters using strong gravitational lensing, it is crucial to determine the positions of a large sample of multiple images from numerous background sources at different redshifts. Likewise, identifying a pure and complete sample of cluster member galaxies is essential for characterising the sub-halo component of the clusters. Additionally, due to degeneracies between lens model parameters, an inaccurate characterisation of the sub-halo mass distribution can introduce biases in determining the other components in \Eq\eqref{eq.: pot_dec}.

Figure \ref{fig:rms} shows the displacements in the sky plane between the observed and model-predicted positions for the multiple images of the $\mathrm{EM_{spec}}$ (red) and $\mathrm{EM_{phot}}$ (blue) models compared to those obtained from the reference lens model by \citealt{Bergamini_M0416_2023} (light grey). We quantified the precision of each model in terms of the root-mean-square separation ($\Delta_{\rm RMS}$) between the observed ($\mathbf{x}^{obs}$) and model-predicted ($\mathbf{x}^{pre}$) positions of the multiple images: 

\begin{equation}
    \Delta_{\rm RMS}=\sqrt{\frac{1}{N_\mathrm{im}^\mathrm{tot}}\sum_{i=1}^{N_\mathrm{im}^\mathrm{tot}}\left\|\mathbf{x}_{i}^\mathrm{pred} - \mathbf{x}_{i}^\mathrm{obs} \right\|^2} \, ,
\end{equation}
where $N_\mathrm{im}^\mathrm{tot}$ is the total number of images in the model.

The $\mathrm{EM_{phot}}$ model is characterised by a $\Delta_{\rm RMS}^\mathrm{EM_{phot}}=\ang{;;0.53}$ in predicting the positions of the multiple images, which is approximately 33\% higher than the other two models. This increased $\Delta_{\rm RMS}$ is attributed to weaker lensing constraints from the less accurate photometric redshift measurements, as opposed to spectroscopic redshifts, for the lensed background sources. In contrast, the $\mathrm{EM_{spec}}$ predicts the positions of the observed multiple images with a $\Delta_{\rm RMS}$ of \ang{;;0.39}, which is about 9\% smaller than for the B23 model (\ang{;;0.43}). This minor discrepancy is due to the differing number of degrees of freedom (DoF, see \Eq 4 in \citealt{Bergamini_2020}) in the two lens models. The $\mathrm{EM_{spec}}$ must reproduce the positions of 31 multiple images from 12 background sources (i.e. about 13\% of the images considered in the B23 model), using 16 free parameters (22 DoF), whereas the B23 model predicts 237 multiple images from 88 sources with 30 free parameters (268 DoF). Therefore, the considerably higher number of DoF in the latter lens model makes it less prone to overfitting, albeit at the cost of a larger $\Delta_{\rm RMS}$ (see \Fig\ref{fig:rms}).

In Figs.\,\ref{fig:kappa} and \ref{fig:magnification}, we compare the convergence and magnification maps obtained from the three lens models. The results are shown on a grid of panels $(i,j)$, where the indices $i,j\in [1,3]$ identify the lens models. Thus, the maps from the $i$-th lens model are displayed along the diagonal ($i=j$) of the figures. The model names are reported at the bottom of each column and on the left side of each row of panels. The panels in the $i-$th row and $j-$th column show the relative differences between the maps of models $j$ and $i$. 

We note that while the convergence maps of the $\mathrm{EM_{spec}}$ and $\mathrm{EM_{phot}}$ models are quite similar, with a median absolute relative difference of $\sim 4$\% (second panel in the bottom row), larger discrepancies exist between these models and the B23 model (first column of panels). This is expected, given the different total mass parametrisations adopted in the \Euclid\ and HST-based lens models. As described in \Sec\ref{sec:lensing}, the large-scale total mass distribution of MACS~J0416 in the $\mathrm{EM_{spec}}$ and $\mathrm{EM_{phot}}$ is parametrised using only two elliptical dPIEs, in contrast to the four included in the B23 model. The two additional haloes in the latter model (a circular one north-east of the northern BCG and a highly elliptical one close to the southern BCG) are clearly identifiable in the panels in the first column of \Fig\ref{fig:kappa}. They correspond to regions of large relative differences between the maps. The models also differ at galaxy scales due to the lower number of cluster galaxies included in the \Euclid-based lens models (125 out of 213 in the B23 model), and due to the absence of the Gaussian prior on the value of the normalisation of the $\sigma_0$--$L$ scaling relation used to model the cluster galaxies (see \Sec\ref{sec:lensing}). In particular, the mass in galaxy-scale haloes in the \Euclid-based lens models is larger than in the B23 model (see \Fig\ref{fig:mass_profile}). Some of the differences between the models are also due to the lack of large-scale haloes describing the hot gas mass distribution in the $\mathrm{EM_{spec}}$ and $\mathrm{EM_{phot}}$ models. 

The \Euclid-based models also exhibit similar magnification patterns, as shown in \Fig\ref{fig:magnification}. However, the magnification is derived from the second spatial derivatives of the lensing potential \citep{Meneghetti_book}. Thus, even small differences between the model mass distributions can lead to large variations in magnification on small scales. These variations are most significant along the critical lines, i.e. the lines where the magnification diverges.  The differences are larger when the \Euclid-based maps are compared to the B23 model. For example, the region along the northern section of the cluster critical lines exhibits large variations between the \Euclid-based and B23 models. The $\mathrm{EM_{spec}}$ and $\mathrm{EM_{phot}}$ models lack constraints in this region. Most of the multiple images detected in the HST data are too faint to be detected by \Euclid, as they originate from distant sources at $z \gtrsim 6$.

Despite these differences, both the $\mathrm{EM_{spec}}$ and $\mathrm{EM_{phot}}$ models can be used to accurately measure the total projected mass profile in the cluster core. In the upper panel of \Fig\ref{fig:mass_profile}, we present the cumulative mass profiles derived from the different lens models. We show profiles for both the total mass distributions and the cluster member (CM) components. We report the ratios between the mass profiles derived from the \Euclid-based and B23 models in the bottom panel. The radius $R$ is measured with respect to the position of the northern BCG (BCG-N). The total cumulative mass profiles agree at the level of $\lesssim 5\%$ in the radial range covered by the strong lensing constraints, as indicated by the vertical segments at the top of the figure. Significant deviations between the \Euclid-based models and the B23 model arise only at distances smaller than $\sim$\,10 kpc from BCG-N. This result is not surprising, as strong lensing is a robust estimator of the total mass within the Einstein radius \citep[i.e. within the lens critical lines; ][]{Meneghetti_2017}. However, as highlighted earlier, the \Euclid-based models measure a larger mass in cluster members compared to the B23 model. Disentangling the large- and small-scale mass components of the cluster requires additional constraints derived from stellar kinematics measurements \citep[e.g.][]{Bergamini_2019,Bergamini_2020}. 

Although not shown in \Fig\ref{fig:mass_profile}, strong lensing alone is unable to constrain the mass profile far outside the region containing the multiple images. However, due to its large field of view and survey strategy, \Euclid\ will measure the shear out to the virial radius and beyond, at least for massive galaxy clusters \citep{Giocoli-EP30,EP-Lesci}. Several studies have demonstrated that combining weak and strong lensing enhances the precision and accuracy of mass profile measurements out to large radii \citep[e.g.][]{Meneghetti_2010}. Moreover, mapping the two-dimensional mass distributions within large regions around the cluster centre allows us to characterise complex mass distributions such as merging clusters and filaments \citep{Bradac_2006,Merten_2011,Merten_2015,diego_2023}.  

\section{\HS\ applications}
\label{sec:applications}

Although this work mainly focuses on the analysis of EWS-like data of galaxy clusters, the \HS\ code is designed to simulate customised \Euclid\ imaging observations. In particular, one can use any kind of HST image as an input and specify a number of parameters (exposure times, PSF models, limiting magnitude, and  S/N) to generate mock images. As an example, \Fig\ref{fig:ggsl} shows simulated \Euclid\ images of two galaxy-galaxy strong lensing systems, as observed in the EWS. The system shown in the left panels of the figure is identified as SDSS~J0946$+$1006, also known as the Jackpot lens, and consists of a double Einstein ring lensed by a galaxy at $z=0.222$ (at the centre of the left images). The inner ring, with a radius of approximately \ang{;;1.4}, has a redshift of $z=0.609$, while the outer ring, with a radius of $\sim$\,\ang{;;2.1}, has $z=2.035$ \citep{Gavazzi_Jackpot}. A third source at $z=5.975$ is also lensed in two additional multiple images \citep{Collett_Jackpot}. The system shown in the right panels is instead a quadruple imaged quasi-stellar object (QSO) at $z=1.693$, lensed by a galaxy at $z=0.455$, known as HE0435$-$1223 \citep{Wisotzki_2002, Bonvin_QSO}.  \Fig\ref{fig:cdfs} shows a simulated preview of a portion of the {\it Chandra} Deep Field South \citep{Giacconi2002} that will be observed in the \Euclid\ Deep Survey (EDS). This simulation assumes an exposure time of $90\,744\,{\rm s}$ and a limiting S/N of 15.9 for a source with an \IE\ magnitude of 26.5, measured within a circular aperture of \ang{;;0.65} radius. These values correspond to the EDS expected requirements. 

Since \HS\ is fully modular, it can be easily generalised to simulate a wide range of imaging data. Although this work focuses on converting HST to \Euclid data, the software can be extended to include multiple input and output instruments, provided that the input images have higher spatial resolution and depth than the simulated output observations. In this context, the new JWST and the future Extremely Large Telescope (ELT) data can be used to generate the simulations. The JWST data can also be used to fully cover the \HE filter as discussed in \Sec\ref{sec:pipeline:combination}.

Another possible application of the \HS\ code, beyond testing the accuracy of \Euclid-based strong lensing models, is the search for transient sources (e.g. supernovae and active galactic nuclei) in previously observed HST (or JWST) fields. In particular, a direct comparison between the upcoming real \Euclid\ data and simulated images allows the identification of transient sources by subtracting the simulated images (before the addition of noise; step d of the simulation pipeline presented in \Sec\,\ref{sec:pipeline} and \Fig\,\ref{fig:pipeline}) from the real observations of the same region. This application is particularly useful for identifying multiple imaged variable sources lensed by galaxy clusters, such as lensed supernovae. The time delays measured between multiple images of these sources can then be used to constrain cosmological parameter values through the time-delay cosmography technique \citep[e.g.][]{Grillo_2020, Treu_2022, Acebron_2023}. 

In addition, \HS\ can be used to provide the training set for CNN-based techniques aimed at identifying cluster members, galaxy-scale lensing systems, and strong lensing features in clusters (e.g. \citealt{Angora_2020, Angora_2023}; Bazzanini et al. in prep.). More generally, mock HST fields with extensive spectroscopic information are crucial for validating the performance of \Euclid\ across a range of legacy science cases (e.g. the search for high-$z$ dropout galaxies and morphological characterisation of galaxies at various redshifts). 

\section{Conclusions}
\label{sec:conclusions}
In this article, we present the \HS\ code developed to create simulated \Euclid images in the \IE, \YE, \JE, and \HE\ bands using real HST observations in the ACS F606W, ACS F814W, WFC3 F105W, WFC3 F125W, and WFC3 F160W filters. The code is written entirely in \texttt{Python} and can be easily customised for a wide range of studies that make use of \Euclid\ images. A high-level interface based on textual input files allows users to access the full functionalities of the code. As a preliminary application, we used \HS\ to simulate EWS data for 27 clusters observed by the HST during the CLASH and HFF surveys (21 CLASH and 6 HFF clusters). The cluster redshifts range from 0.19 to 0.89. 

By using the simulated \Euclid\ images of the galaxy cluster MACS~J0416, we tested the possibility of developing high-precision strong lensing models of galaxy clusters based on \Euclid\ data. Our results demonstrate that the \Euclid-based lens models are sufficiently accurate to yield precise estimates of the total mass within the cluster critical lines. 

The precision and accuracy achievable with \Euclid-based lens models rely on the identification of multiple images and cluster members. Both types of sources serve as inputs for constructing the lens models. In this context, redshift measurements from spectroscopic follow-up campaigns will be essential. However, machine and deep learning techniques can automate the search for these sources. Given the paucity of known lenses, especially on the scale of galaxy clusters, training deep learning models requires realistic and sophisticated image simulations, which \HS\ can deliver. For instance, strongly lensed galaxies can be injected into Euclidised HST images to construct a training set (Angora et al. in prep.; Bazzanini et al. in prep.).

Despite our work mainly focusing on analysing EWS-like data of galaxy clusters, the \HS\ code is designed to simulate any kind of \Euclid\ imaging data. 

\begin{acknowledgements}
      We acknowledge financial support through grants PRIN-MIUR 2017WSCC32 and 2020SKSTHZ. PB acknowledges financial support from ASI through the agreement ASI-INAF n. 2018-29-HH.0. MM acknowledges support from the Italian Space Agency (ASI) through contract ``Euclid - Phase E'' and ``Euclid - Phase D''. LB is indebted to the communities behind the multiple free, libre, and open-source software packages on which we all depend on. \AckEC
      This work uses the following software packages:
    \href{https://github.com/astropy/astropy}{\texttt{Astropy}}
    \citep{astropy1, astropy2},
    \href{https://github.com/matplotlib/matplotlib}{\texttt{matplotlib}}
    \citep{matplotlib},
    \href{https://github.com/numpy/numpy}{\texttt{NumPy}}
    \citep{numpy1, numpy2},
    \href{https://www.python.org/}{\texttt{Python}}
    \citep{python},
    \href{https://github.com/scipy/scipy}{\texttt{Scipy}}
    \citep{scipy},
    \href{https://acstools.readthedocs.io/en/latest/}{\texttt{acstools}}
    \citep{acstools},
    \href{https://reproject.readthedocs.io/en/stable/}{\texttt{reproject}},    \href{https://docs.python.org/3/library/argparse.html}{\texttt{argparse}},    \href{https://docs.python.org/3/library/re.html}{\texttt{re}},    \href{https://docs.python.org/3/library/pickle.html}{\texttt{pickle}},    \href{https://docs.python.org/3/library/os.html}{\texttt{os}},    \href{https://photutils.readthedocs.io/en/stable/}{\texttt{photutils}}
    \citep{photutils}.
\end{acknowledgements}

\bibliographystyle{aa}
\bibliography{bibliography, Euclid}

\begin{thebibliography}{79}
\expandafter\ifx\csname natexlab\endcsname\relax\def\natexlab#1{#1}\fi

\bibitem[{{Acebron} {et~al.}(2023){Acebron}, {Schuldt}, {Grillo}, {Bergamini}, {Granata}, {Me{\v{s}}tri{\'c}}, {Caminha}, {Meneghetti}, {Mercurio}, {Rosati}, {Suyu}, \& {Vanzella}}]{Acebron_2023}
{Acebron}, A., {Schuldt}, S., {Grillo}, C., {et~al.} 2023, \aap, 680, L9

\bibitem[{{Angora} {et~al.}(2020){Angora}, {Rosati}, {Brescia}, {Mercurio}, {Grillo}, {Caminha}, {Meneghetti}, {Nonino}, {Vanzella}, {Bergamini}, {Biviano}, \& {Lombardi}}]{Angora_2020}
{Angora}, G., {Rosati}, P., {Brescia}, M., {et~al.} 2020, \aap, 643, A177

\bibitem[{{Angora} {et~al.}(2023){Angora}, {Rosati}, {Meneghetti}, {Brescia}, {Mercurio}, {Grillo}, {Bergamini}, {Acebron}, {Caminha}, {Nonino}, {Tortorelli}, {Bazzanini}, \& {Vanzella}}]{Angora_2023}
{Angora}, G., {Rosati}, P., {Meneghetti}, M., {et~al.} 2023, \aap, 676, A40

\bibitem[{{Astropy Collaboration: Robitaille} {et~al.}(2013){Astropy Collaboration: Robitaille}, {Tollerud}, {Greenfield}, {Droettboom}, {Bray}, {Aldcroft}, {Davis}, {Ginsburg}, {Price-Whelan}, {Kerzendorf}, {Conley}, {Crighton}, {Barbary}, {Muna}, {Ferguson}, {Grollier}, {Parikh}, {Nair}, {Unther}, {Deil}, {Woillez}, {Conseil}, {Kramer}, {Turner}, {Singer}, {Fox}, {Weaver}, {Zabalza}, {Edwards}, {Azalee Bostroem}, {Burke}, {Casey}, {Crawford}, {Dencheva}, {Ely}, {Jenness}, {Labrie}, {Lim}, {Pierfederici}, {Pontzen}, {Ptak}, {Refsdal}, {Servillat}, \& {Streicher}}]{astropy1}
{Astropy Collaboration: Robitaille}, T.~P., {Tollerud}, E.~J., {Greenfield}, P., {et~al.} 2013, \aap, 558, A33

\bibitem[{{Bacon} {et~al.}(2012){Bacon}, {Accardo}, {Adjali}, {Anwand}, {Bauer}, {Blaizot}, {Boudon}, {Brinchmann}, {Brotons}, {Caillier}, {Capoani}, {Carollo}, {Comin}, {Contini}, {Cumani}, {Daguis}, {Deiries}, {Delabre}, {Dreizler}, {Dubois}, {Dupieux}, {Dupuy}, {Emsellem}, {Fleischmann}, {Fran{\c{c}}ois}, {Gallou}, {Gharsa}, {Girard}, {Glindemann}, {Guiderdoni}, {Hahn}, {Hansali}, {Hofmann}, {Jarno}, {Kelz}, {Kiekebusch}, {Knudstrup}, {Koehler}, {Kollatschny}, {Kosmalski}, {Laurent}, {Le Floch}, {Lilly}, {Lizon {\`a} L'Allemand}, {Loupias}, {Manescau}, {Monstein}, {Nicklas}, {Niemeyer}, {Olaya}, {Palsa}, {Par{\`e}s}, {Pasquini}, {P{\'e}contal-Rousset}, {Pello}, {Petit}, {Piqueras}, {Popow}, {Reiss}, {Remillieux}, {Renault}, {Rhode}, {Richard}, {Roth}, {Rupprecht}, {Schaye}, {Slezak}, {Soucail}, {Steinmetz}, {Streicher}, {Stuik}, {Valentin}, {Vernet}, {Weilbacher}, {Wisotzki}, {Yerle}, \& {Zins}}]{Bacon_MUSE}
{Bacon}, R., {Accardo}, M., {Adjali}, L., {et~al.} 2012, The Messenger, 147, 4

\bibitem[{{Bartelmann} {et~al.}(2013){Bartelmann}, {Limousin}, {Meneghetti}, \& {Schmidt}}]{Bartelmann_2013}
{Bartelmann}, M., {Limousin}, M., {Meneghetti}, M., \& {Schmidt}, R. 2013, \ssr, 177, 3

\bibitem[{{Bayliss} {et~al.}(2011){Bayliss}, {Hennawi}, {Gladders}, {Koester}, {Sharon}, {Dahle}, \& {Oguri}}]{Bayliss_2011}
{Bayliss}, M.~B., {Hennawi}, J.~F., {Gladders}, M.~D., {et~al.} 2011, \apjs, 193, 8

\bibitem[{{Bergamini} {et~al.}(2023){Bergamini}, {Grillo}, {Rosati}, {Vanzella}, {Me{\v{s}}tri{\'c}}, {Mercurio}, {Acebron}, {Caminha}, {Granata}, {Meneghetti}, {Angora}, \& {Nonino}}]{Bergamini_M0416_2023}
{Bergamini}, P., {Grillo}, C., {Rosati}, P., {et~al.} 2023, \aap, 674, A79

\bibitem[{{Bergamini} {et~al.}(2019){Bergamini}, {Rosati}, {Mercurio}, {Grillo}, {Caminha}, {Meneghetti}, {Agnello}, {Biviano}, {Calura}, {Giocoli}, {Lombardi}, {Rodighiero}, \& {Vanzella}}]{Bergamini_2019}
{Bergamini}, P., {Rosati}, P., {Mercurio}, A., {et~al.} 2019, \aap, 631, A130

\bibitem[{{Bergamini} {et~al.}(2021){Bergamini}, {Rosati}, {Vanzella}, {Caminha}, {Grillo}, {Mercurio}, {Meneghetti}, {Angora}, {Calura}, {Nonino}, \& {Tozzi}}]{Bergamini_2020}
{Bergamini}, P., {Rosati}, P., {Vanzella}, E., {et~al.} 2021, \aap, 645, A140

\bibitem[{{Boldrin} {et~al.}(2012){Boldrin}, {Giocoli}, {Meneghetti}, \& {Moscardini}}]{Boldrin_2012}
{Boldrin}, M., {Giocoli}, C., {Meneghetti}, M., \& {Moscardini}, L. 2012, \mnras, 427, 3134

\bibitem[{{Boldrin} {et~al.}(2016){Boldrin}, {Giocoli}, {Meneghetti}, {Moscardini}, {Tormen}, \& {Biviano}}]{Boldrin_2016}
{Boldrin}, M., {Giocoli}, C., {Meneghetti}, M., {et~al.} 2016, \mnras, 457, 2738

\bibitem[{{Bonamigo} {et~al.}(2018){Bonamigo}, {Grillo}, {Ettori}, {Caminha}, {Rosati}, {Mercurio}, {Munari}, {Annunziatella}, {Balestra}, \& {Lombardi}}]{Bonamigo_2018}
{Bonamigo}, M., {Grillo}, C., {Ettori}, S., {et~al.} 2018, \apj, 864, 98

\bibitem[{{Bonvin} {et~al.}(2017){Bonvin}, {Courbin}, {Suyu}, {Marshall}, {Rusu}, {Sluse}, {Tewes}, {Wong}, {Collett}, {Fassnacht}, {Treu}, {Auger}, {Hilbert}, {Koopmans}, {Meylan}, {Rumbaugh}, {Sonnenfeld}, \& {Spiniello}}]{Bonvin_QSO}
{Bonvin}, V., {Courbin}, F., {Suyu}, S.~H., {et~al.} 2017, \mnras, 465, 4914

\bibitem[{{Brada{\v{c}}} {et~al.}(2006){Brada{\v{c}}}, {Clowe}, {Gonzalez}, {Marshall}, {Forman}, {Jones}, {Markevitch}, {Randall}, {Schrabback}, \& {Zaritsky}}]{Bradac_2006}
{Brada{\v{c}}}, M., {Clowe}, D., {Gonzalez}, A.~H., {et~al.} 2006, \apj, 652, 937

\bibitem[{{Brada{\v{c}}} {et~al.}(2005){Brada{\v{c}}}, {Schneider}, {Lombardi}, \& {Erben}}]{Bradac_2004}
{Brada{\v{c}}}, M., {Schneider}, P., {Lombardi}, M., \& {Erben}, T. 2005, \aap, 437, 39

\bibitem[{Bradley {et~al.}(2023)Bradley, Sip{\H o}cz, Robitaille, Tollerud, Vin{\'{\i}}cius, Deil, Barbary, Wilson, Busko, Donath, G{\"u}nther, Cara, Lim, Me{\ss}linger, Conseil, Bostroem, Droettboom, Bray, Bratholm, Barentsen, Craig, Rathi, Pascual, Perren, Georgiev, de~Val-Borro, Kerzendorf, Bach, Quint, \& Souchereau}]{photutils}
Bradley, L., Sip{\H o}cz, B., Robitaille, T., {et~al.} 2023, astropy/photutils: 1.8.0

\bibitem[{{Capak} {et~al.}(2007){Capak}, {Aussel}, {Ajiki}, {McCracken}, {Mobasher}, {Scoville}, {Shopbell}, {Taniguchi}, {Thompson}, {Tribiano}, {Sasaki}, {Blain}, {Brusa}, {Carilli}, {Comastri}, {Carollo}, {Cassata}, {Colbert}, {Ellis}, {Elvis}, {Giavalisco}, {Green}, {Guzzo}, {Hasinger}, {Ilbert}, {Impey}, {Jahnke}, {Kartaltepe}, {Kneib}, {Koda}, {Koekemoer}, {Komiyama}, {Leauthaud}, {Le Fevre}, {Lilly}, {Liu}, {Massey}, {Miyazaki}, {Murayama}, {Nagao}, {Peacock}, {Pickles}, {Porciani}, {Renzini}, {Rhodes}, {Rich}, {Salvato}, {Sanders}, {Scarlata}, {Schiminovich}, {Schinnerer}, {Scodeggio}, {Sheth}, {Shioya}, {Tasca}, {Taylor}, {Yan}, \& {Zamorani}}]{Capak2007}
{Capak}, P., {Aussel}, H., {Ajiki}, M., {et~al.} 2007, \apjs, 172, 99

\bibitem[{{Coe} {et~al.}(2008){Coe}, {Fuselier}, {Ben{\'{\i}}tez}, {Broadhurst}, {Frye}, \& {Ford}}]{Coe08}
{Coe}, D., {Fuselier}, E., {Ben{\'{\i}}tez}, N., {et~al.} 2008, \apj, 681, 814

\bibitem[{{Collett} \& {Smith}(2020)}]{Collett_Jackpot}
{Collett}, T.~E. \& {Smith}, R.~J. 2020, \mnras, 497, 1654

\bibitem[{{Cropper} {et~al.}(2016){Cropper}, {Pottinger}, {Niemi}, {Azzollini}, {Denniston}, {Szafraniec}, {Awan}, {Mellier}, {Berthe}, {Martignac}, {Cara}, {Di Giorgio}, {Sciortino}, {Bozzo}, {Genolet}, {Cole}, {Philippon}, {Hailey}, {Hunt}, {Swindells}, {Holland}, {Gow}, {Murray}, {Hall}, {Skottfelt}, {Amiaux}, {Laureijs}, {Racca}, {Salvignol}, {Short}, {Lorenzo Alvarez}, {Kitching}, {Hoekstra}, {Massey}, \& {Israel}}]{Cropper_2016}
{Cropper}, M., {Pottinger}, S., {Niemi}, S., {et~al.} 2016, in Society of Photo-Optical Instrumentation Engineers (SPIE) Conference Series, Vol. 9904, Space Telescopes and Instrumentation 2016: Optical, Infrared, and Millimeter Wave, ed. H.~A. {MacEwen}, G.~G. {Fazio}, M.~{Lystrup}, N.~{Batalha}, N.~{Siegler}, \& E.~C. {Tong}, 99040Q

\bibitem[{{Diego} {et~al.}(2023){Diego}, {Meena}, {Adams}, {Broadhurst}, {Dai}, {Coe}, {Frye}, {Kelly}, {Koekemoer}, {Pascale}, {Willner}, {Zackrisson}, {Zitrin}, {Windhorst}, {Cohen}, {Jansen}, {Summers}, {Tompkins}, {Conselice}, {Driver}, {Yan}, {Grogin}, {Marshall}, {Pirzkal}, {Robotham}, {Ryan}, {Willmer}, {Bradley}, {Caminha}, {Caputi}, {Carleton}, \& {Kamieneski}}]{diego_2023}
{Diego}, J.~M., {Meena}, A.~K., {Adams}, N.~J., {et~al.} 2023, \aap, 672, A3

\bibitem[{{Diego} {et~al.}(2005){Diego}, {Protopapas}, {Sandvik}, \& {Tegmark}}]{Diego2005}
{Diego}, J.~M., {Protopapas}, P., {Sandvik}, H.~B., \& {Tegmark}, M. 2005, \mnras, 360, 477

\bibitem[{{Ebeling} {et~al.}(2007){Ebeling}, {Barrett}, {Donovan}, {Ma}, {Edge}, \& {van Speybroeck}}]{Ebeling_2007}
{Ebeling}, H., {Barrett}, E., {Donovan}, D., {et~al.} 2007, \apjl, 661, L33

\bibitem[{{Ebeling} {et~al.}(2001){Ebeling}, {Edge}, \& {Henry}}]{Ebeling_2001}
{Ebeling}, H., {Edge}, A.~C., \& {Henry}, J.~P. 2001, \apj, 553, 668

\bibitem[{{Ebeling} {et~al.}(2010){Ebeling}, {Edge}, {Mantz}, {Barrett}, {Henry}, {Ma}, \& {van Speybroeck}}]{Ebeling_2010}
{Ebeling}, H., {Edge}, A.~C., {Mantz}, A., {et~al.} 2010, \mnras, 407, 83

\bibitem[{{El{\'\i}asd{\'o}ttir} {et~al.}(2007){El{\'\i}asd{\'o}ttir}, {Limousin}, {Richard}, {Hjorth}, {Kneib}, {Natarajan}, {Pedersen}, {Jullo}, \& {Paraficz}}]{Eliasdottir_lenstool}
{El{\'\i}asd{\'o}ttir}, {\'A}., {Limousin}, M., {Richard}, J., {et~al.} 2007, arXiv e-prints, arXiv:0710.5636

\bibitem[{{Euclid Collaboration: Desprez} {et~al.}(2020){Euclid Collaboration: Desprez}, {Paltani}, {Coupon}, {et~al.}}]{Desprez-EP10}
{Euclid Collaboration: Desprez}, G., {Paltani}, S., {Coupon}, J., {et~al.} 2020, \aap, 644, A31

\bibitem[{{Euclid Collaboration: Giocoli} {et~al.}(2024){Euclid Collaboration: Giocoli}, {Meneghetti}, {Rasia}, {et~al.}}]{Giocoli-EP30}
{Euclid Collaboration: Giocoli}, C., {Meneghetti}, M., {Rasia}, E., {et~al.} 2024, \aap, 681, A67

\bibitem[{{Euclid Collaboration: Lesci} {et~al.}(2024){Euclid Collaboration: Lesci}, {Sereno}, {Radovich}, {et~al.}}]{EP-Lesci}
{Euclid Collaboration: Lesci}, G.~F., {Sereno}, M., {Radovich}, M., {et~al.} 2024, \aap, 684, A139

\bibitem[{{Euclid Collaboration: Mellier} {et~al.}(2025){Euclid Collaboration: Mellier}, {Abdurro'uf}, {Acevedo~Barroso}, {et~al.}}]{EuclidSkyOverview}
{Euclid Collaboration: Mellier}, Y., {Abdurro'uf}, {Acevedo~Barroso}, J., {et~al.} 2025, A\&A, 697, A1

\bibitem[{{Euclid Collaboration: Paltani} {et~al.}(2024){Euclid Collaboration: Paltani}, {Coupon}, {Hartley}, {et~al.}}]{EP-Paltani}
{Euclid Collaboration: Paltani}, S., {Coupon}, J., {Hartley}, W.~G., {et~al.} 2024, \aap, 681, A66

\bibitem[{{Euclid Collaboration: Scaramella} {et~al.}(2022){Euclid Collaboration: Scaramella}, {Amiaux}, {Mellier}, {et~al.}}]{Scaramella-EP1}
{Euclid Collaboration: Scaramella}, R., {Amiaux}, J., {Mellier}, Y., {et~al.} 2022, \aap, 662, A112

\bibitem[{{Euclid Collaboration: Schirmer} {et~al.}(2022){Euclid Collaboration: Schirmer}, {Jahnke}, {Seidel}, {et~al.}}]{Schirmer-EP18}
{Euclid Collaboration: Schirmer}, M., {Jahnke}, K., {Seidel}, G., {et~al.} 2022, \aap, 662, A92

\bibitem[{{Gavazzi} {et~al.}(2008){Gavazzi}, {Treu}, {Koopmans}, {Bolton}, {Moustakas}, {Burles}, \& {Marshall}}]{Gavazzi_Jackpot}
{Gavazzi}, R., {Treu}, T., {Koopmans}, L. V.~E., {et~al.} 2008, \apj, 677, 1046

\bibitem[{{Giacconi} {et~al.}(2002){Giacconi}, {Zirm}, {Wang}, {Rosati}, {Nonino}, {Tozzi}, {Gilli}, {Mainieri}, {Hasinger}, {Kewley}, {Bergeron}, {Borgani}, {Gilmozzi}, {Grogin}, {Koekemoer}, {Schreier}, {Zheng}, \& {Norman}}]{Giacconi2002}
{Giacconi}, R., {Zirm}, A., {Wang}, J., {et~al.} 2002, \apjs, 139, 369

\bibitem[{{Granata} {et~al.}(2022){Granata}, {Mercurio}, {Grillo}, {Tortorelli}, {Bergamini}, {Meneghetti}, {Rosati}, {Caminha}, \& {Nonino}}]{Granata_2021}
{Granata}, G., {Mercurio}, A., {Grillo}, C., {et~al.} 2022, \aap, 659, A24

\bibitem[{{Grillo} {et~al.}(2020){Grillo}, {Rosati}, {Suyu}, {Caminha}, {Mercurio}, \& {Halkola}}]{Grillo_2020}
{Grillo}, C., {Rosati}, P., {Suyu}, S.~H., {et~al.} 2020, \apj, 898, 87

\bibitem[{Harris {et~al.}(2020)Harris, Millman, van~der Walt, Gommers, Virtanen, Cournapeau, Wieser, Taylor, Berg, Smith, Kern, Picus, Hoyer, van Kerkwijk, Brett, Haldane, Fernández~del Río, Wiebe, Peterson, Gérard-Marchant, Sheppard, Reddy, Weckesser, Abbasi, Gohlke, \& Oliphant}]{numpy2}
Harris, C.~R., Millman, K.~J., van~der Walt, S.~J., {et~al.} 2020, Nature, 585, 357

\bibitem[{{Hogg} {et~al.}(2002){Hogg}, {Baldry}, {Blanton}, \& {Eisenstein}}]{Hogg_2002}
{Hogg}, D.~W., {Baldry}, I.~K., {Blanton}, M.~R., \& {Eisenstein}, D.~J. 2002, arXiv e-prints, arXiv:astro-ph/0210394

\bibitem[{Hunter(2007)}]{matplotlib}
Hunter, J.~D. 2007, Computing in Science \& Engineering, 9, 90

\bibitem[{{Johnson} \& {Sharon}(2016)}]{Johnson_2016}
{Johnson}, T.~L. \& {Sharon}, K. 2016, \apj, 832, 82

\bibitem[{{Jullo} \& {Kneib}(2009)}]{Jullo_Kneib_lenstool}
{Jullo}, E. \& {Kneib}, J.-P. 2009, \mnras, 395, 1319

\bibitem[{{Jullo} {et~al.}(2007){Jullo}, {Kneib}, {Limousin}, {El{\'{\i}}asd{\'o}ttir}, {Marshall}, \& {Verdugo}}]{Jullo_lenstool}
{Jullo}, E., {Kneib}, J.-P., {Limousin}, M., {et~al.} 2007, New Journal of Physics, 9, 447

\bibitem[{Kneib {et~al.}(1993)Kneib, Mellier, Fort, \& Mathez}]{Kneib_1993}
Kneib, J., Mellier, Y., Fort, B., \& Mathez, G. 1993, A\&A, 273, 367

\bibitem[{{Kneib} \& {Natarajan}(2011)}]{Kneib_2011}
{Kneib}, J.-P. \& {Natarajan}, P. 2011, \aapr, 19, 47

\bibitem[{{Lam} {et~al.}(2014){Lam}, {Broadhurst}, {Diego}, {Lim}, {Coe}, {Ford}, \& {Zheng}}]{Lam2014}
{Lam}, D., {Broadhurst}, T., {Diego}, J.~M., {et~al.} 2014, \apj, 797, 98

\bibitem[{{Laureijs} {et~al.}(2011){Laureijs}, {Amiaux}, {Arduini}, {Augu{\`e}res}, {Brinchmann}, {Cole}, {Cropper}, {Dabin}, {Duvet}, {Ealet}, {Garilli}, {Gondoin}, {Guzzo}, {Hoar}, {Hoekstra}, {Holmes}, {Kitching}, {Maciaszek}, {Mellier}, {Pasian}, {Percival}, {Rhodes}, {Saavedra Criado}, {Sauvage}, {Scaramella}, {Valenziano}, {Warren}, {Bender}, {Castander}, {Cimatti}, {Le F{\`e}vre}, {Kurki-Suonio}, {Levi}, {Lilje}, {Meylan}, {Nichol}, {Pedersen}, {Popa}, {Rebolo Lopez}, {Rix}, {Rottgering}, {Zeilinger}, {Grupp}, {Hudelot}, {Massey}, {Meneghetti}, {Miller}, {Paltani}, {Paulin-Henriksson}, {Pires}, {Saxton}, {Schrabback}, {Seidel}, {Walsh}, {Aghanim}, {Amendola}, {Bartlett}, {Baccigalupi}, {Beaulieu}, {Benabed}, {Cuby}, {Elbaz}, {Fosalba}, {Gavazzi}, {Helmi}, {Hook}, {Irwin}, {Kneib}, {Kunz}, {Mannucci}, {Moscardini}, {Tao}, {Teyssier}, {Weller}, {Zamorani}, {Zapatero Osorio}, {Boulade}, {Foumond}, {Di Giorgio}, {Guttridge}, {James}, {Kemp}, {Martignac}, {Spencer}, {Walton}, {Bl{\"u}mchen}, {Bonoli},
  {Bortoletto}, {Cerna}, {Corcione}, {Fabron}, {Jahnke}, {Ligori}, {Madrid}, {Martin}, {Morgante}, {Pamplona}, {Prieto}, {Riva}, {Toledo}, {Trifoglio}, {Zerbi}, {Abdalla}, {Douspis}, {Grenet}, {Borgani}, {Bouwens}, {Courbin}, {Delouis}, {Dubath}, {Fontana}, {Frailis}, {Grazian}, {Koppenh{\"o}fer}, {Mansutti}, {Melchior}, {Mignoli}, {Mohr}, {Neissner}, {Noddle}, {Poncet}, {Scodeggio}, {Serrano}, {Shane}, {Starck}, {Surace}, {Taylor}, {Verdoes-Kleijn}, {Vuerli}, {Williams}, {Zacchei}, {Altieri}, {Escudero Sanz}, {Kohley}, {Oosterbroek}, {Astier}, {Bacon}, {Bardelli}, {Baugh}, {Bellagamba}, {Benoist}, {Bianchi}, {Biviano}, {Branchini}, {Carbone}, {Cardone}, {Clements}, {Colombi}, {Conselice}, {Cresci}, {Deacon}, {Dunlop}, {Fedeli}, {Fontanot}, {Franzetti}, {Giocoli}, {Garcia-Bellido}, {Gow}, {Heavens}, {Hewett}, {Heymans}, {Holland}, {Huang}, {Ilbert}, {Joachimi}, {Jennins}, {Kerins}, {Kiessling}, {Kirk}, {Kotak}, {Krause}, {Lahav}, {van Leeuwen}, {Lesgourgues}, {Lombardi}, {Magliocchetti}, {Maguire},
  {Majerotto}, {Maoli}, {Marulli}, {Maurogordato}, {McCracken}, {McLure}, {Melchiorri}, {Merson}, {Moresco}, {Nonino}, {Norberg}, {Peacock}, {Pello}, {Penny}, {Pettorino}, {Di Porto}, {Pozzetti}, {Quercellini}, {Radovich}, {Rassat}, {Roche}, {Ronayette}, {Rossetti}, {Sartoris}, {Schneider}, {Semboloni}, {Serjeant}, {Simpson}, {Skordis}, {Smadja}, {Smartt}, {Spano}, {Spiro}, {Sullivan}, {Tilquin}, {Trotta}, {Verde}, {Wang}, {Williger}, {Zhao}, {Zoubian}, \& {Zucca}}]{Laureijs11}
{Laureijs}, R., {Amiaux}, J., {Arduini}, S., {et~al.} 2011, ESA/SRE(2011)12, arXiv:1110.3193

\bibitem[{{Liesenborgs} {et~al.}(2006){Liesenborgs}, {De Rijcke}, \& {Dejonghe}}]{lie06}
{Liesenborgs}, J., {De Rijcke}, S., \& {Dejonghe}, H. 2006, \mnras, 367, 1209

\bibitem[{{Lim} {et~al.}(2020){Lim}, {Davis}, {Hack}, {Grogin}, {Ogaz}, {Ubeda}, {Cara}, {Borncamp}, \& {Miles}}]{acstools}
{Lim}, P.~L., {Davis}, M., {Hack}, W., {et~al.} 2020, {ACStools: Python tools for Hubble Space Telescope Advanced Camera for Surveys data}, Astrophysics Source Code Library, record ascl:2011.024

\bibitem[{{Limousin} {et~al.}(2005){Limousin}, {Kneib}, \& {Natarajan}}]{Limousin_lenstool}
{Limousin}, M., {Kneib}, J.-P., \& {Natarajan}, P. 2005, \mnras, 356, 309

\bibitem[{{Lotz} {et~al.}(2014){Lotz}, {Mountain}, {Grogin}, {Koekemoer}, {Capak}, {Mack}, {Coe}, {Barker}, {Adler}, {Avila}, {Anderson}, {Casertano}, {Christian}, {Gonzaga}, {Ferguson}, {Fruchter}, {Jenkner}, {Jordan}, {Hammer}, {Hilbert}, {Lawton}, {Lee}, {Lucas}, {MacKenty}, {Mutchler}, {Ogaz}, {Reid}, {Royle}, {Robberto}, {Sembach}, {Smith}, {Sokol}, {Surace}, {Taylor}, {Tumlinson}, {Viana}, {Williams}, \& {Workman}}]{Lotz_2014HFF}
{Lotz}, J., {Mountain}, M., {Grogin}, N.~A., {et~al.} 2014, in American Astronomical Society Meeting Abstracts, Vol. 223, American Astronomical Society Meeting Abstracts \#223, 254.01

\bibitem[{{Lotz} {et~al.}(2017){Lotz}, {Koekemoer}, {Coe}, {Grogin}, {Capak}, {Mack}, {Anderson}, {Avila}, {Barker}, {Borncamp}, {Brammer}, {Durbin}, {Gunning}, {Hilbert}, {Jenkner}, {Khandrika}, {Levay}, {Lucas}, {MacKenty}, {Ogaz}, {Porterfield}, {Reid}, {Robberto}, {Royle}, {Smith}, {Storrie-Lombardi}, {Sunnquist}, {Surace}, {Taylor}, {Williams}, {Bullock}, {Dickinson}, {Finkelstein}, {Natarajan}, {Richard}, {Robertson}, {Tumlinson}, {Zitrin}, {Flanagan}, {Sembach}, {Soifer}, \& {Mountain}}]{Lotz_2017HFF}
{Lotz}, J.~M., {Koekemoer}, A., {Coe}, D., {et~al.} 2017, \apj, 837, 97

\bibitem[{{Maciaszek} {et~al.}(2022){Maciaszek}, {Ealet}, {Gillard}, {Jahnke}, {Barbier}, {Prieto}, {Bon}, {Bonnefoi}, {Caillat}, {Carle}, {Costille}, {Ducret}, {Fabron}, {Foulon}, {Gimenez}, {Grassi}, {Jaquet}, {Le Mignant}, {Martin}, {Pamplona}, {Sanchez}, {Cl{\'e}mens}, {Caillat}, {Niclas}, {Secroun}, {Kubik}, {Ferriol}, {Berthe}, {Barri{\`e}re}, {Fontignie}, {Valenziano}, {Auricchio}, {Battaglia}, {De Rosa}, {Farinelli}, {Franceschi}, {Medinaceli}, {Morgante}, {Sortino}, {Trifoglio}, {Corcione}, {Capobianco}, {Ligori}, {Dusini}, {Borsato}, {Dal Corso}, {Laudisio}, {Sirignano}, {Stanco}, {Ventura}, {Patrizii}, {Chiarusi}, {Fornari}, {Giacomini}, {Margiotta}, {Mauri}, {Pasqualini}, {Sirri}, {Spurio}, {Tenti}, {Travaglini}, {Bonoli}, {Bortoletto}, {Balestra}, {Dalessandro}, {Grupp}, {Penka}, {Steinwagner}, {Hormuth}, {Schirmer}, {Seidel}, {Padilla}, {Casas}, {Lloro}, {Toledo-Moreo}, {Gomez}, {Colodro-Conde}, {Liz{\'a}n}, {Diaz}, {Lilje}, {Andersen}, {Andersen}, {S{\o}rensen}, {Hornstrup}, {Jessen}, {Thizy},
  {Holmes}, {Pniel}, {Jhabvala}, {Pravdo}, {Seiffert}, {Waczynski}, {Laureij}, {Racca}, {Salvignol}, {Boenke}, {Strada}, \& {Mellier}}]{Maciaszek22}
{Maciaszek}, T., {Ealet}, A., {Gillard}, W., {et~al.} 2022, in Society of Photo-Optical Instrumentation Engineers (SPIE) Conference Series, Vol. 12180, Space Telescopes and Instrumentation 2022: Optical, Infrared, and Millimeter Wave, ed. L.~E. {Coyle}, S.~{Matsuura}, \& M.~D. {Perrin}, arXiv:2210.10112

\bibitem[{Meneghetti(2021)}]{Meneghetti_book}
Meneghetti, M. 2021, Introduction to Gravitational Lensing: With Python Examples, Lecture Notes in Physics (Springer International Publishing)

\bibitem[{{Meneghetti} {et~al.}(2013){Meneghetti}, {Bartelmann}, {Dahle}, \& {Limousin}}]{Meneghetti_2013}
{Meneghetti}, M., {Bartelmann}, M., {Dahle}, H., \& {Limousin}, M. 2013, \ssr, 177, 31

\bibitem[{{Meneghetti} {et~al.}(2023){Meneghetti}, {Cui}, {Rasia}, {Yepes}, {Acebron}, {Angora}, {Bergamini}, {Borgani}, {Calura}, {Despali}, {Giocoli}, {Granata}, {Grillo}, {Knebe}, {Macci{\`o}}, {Mercurio}, {Moscardini}, {Natarajan}, {Ragagnin}, {Rosati}, \& {Vanzella}}]{Meneghetti_2023}
{Meneghetti}, M., {Cui}, W., {Rasia}, E., {et~al.} 2023, \aap, 678, L2

\bibitem[{{Meneghetti} {et~al.}(2020){Meneghetti}, {Davoli}, {Bergamini}, {Rosati}, {Natarajan}, {Giocoli}, {Caminha}, {Metcalf}, {Rasia}, {Borgani}, {Calura}, {Grillo}, {Mercurio}, \& {Vanzella}}]{Meneghetti_2020}
{Meneghetti}, M., {Davoli}, G., {Bergamini}, P., {et~al.} 2020, Science, 369, 1347

\bibitem[{{Meneghetti} {et~al.}(2017){Meneghetti}, {Natarajan}, {Coe}, {Contini}, {De Lucia}, {Giocoli}, {Acebron}, {Borgani}, {Bradac}, {Diego}, {Hoag}, {Ishigaki}, {Johnson}, {Jullo}, {Kawamata}, {Lam}, {Limousin}, {Liesenborgs}, {Oguri}, {Sebesta}, {Sharon}, {Williams}, \& {Zitrin}}]{Meneghetti_2017}
{Meneghetti}, M., {Natarajan}, P., {Coe}, D., {et~al.} 2017, \mnras, 472, 3177

\bibitem[{{Meneghetti} {et~al.}(2022){Meneghetti}, {Ragagnin}, {Borgani}, {Calura}, {Despali}, {Giocoli}, {Granato}, {Grillo}, {Moscardini}, {Rasia}, {Rosati}, {Angora}, {Bassini}, {Bergamini}, {Caminha}, {Granata}, {Mercurio}, {Metcalf}, {Natarajan}, {Nonino}, {Pignataro}, {Ragone-Figueroa}, {Vanzella}, {Acebron}, {Dolag}, {Murante}, {Taffoni}, {Tornatore}, {Tortorelli}, \& {Valentini}}]{Meneghetti_2022}
{Meneghetti}, M., {Ragagnin}, A., {Borgani}, S., {et~al.} 2022, \aap, 668, A188

\bibitem[{{Meneghetti} {et~al.}(2010){Meneghetti}, {Rasia}, {Merten}, {Bellagamba}, {Ettori}, {Mazzotta}, {Dolag}, \& {Marri}}]{Meneghetti_2010}
{Meneghetti}, M., {Rasia}, E., {Merten}, J., {et~al.} 2010, \aap, 514, A93

\bibitem[{{Merten} {et~al.}(2011){Merten}, {Coe}, {Dupke}, {Massey}, {Zitrin}, {Cypriano}, {Okabe}, {Frye}, {Braglia}, {Jim{\'e}nez-Teja}, {Ben{\'\i}tez}, {Broadhurst}, {Rhodes}, {Meneghetti}, {Moustakas}, {Sodr{\'e}}, {Krick}, \& {Bregman}}]{Merten_2011}
{Merten}, J., {Coe}, D., {Dupke}, R., {et~al.} 2011, \mnras, 417, 333

\bibitem[{{Merten} {et~al.}(2015){Merten}, {Meneghetti}, {Postman}, {Umetsu}, {Zitrin}, {Medezinski}, {Nonino}, {Koekemoer}, {Melchior}, {Gruen}, {Moustakas}, {Bartelmann}, {Host}, {Donahue}, {Coe}, {Molino}, {Jouvel}, {Monna}, {Seitz}, {Czakon}, {Lemze}, {Sayers}, {Balestra}, {Rosati}, {Ben{\'{\i}}tez}, {Biviano}, {Bouwens}, {Bradley}, {Broadhurst}, {Carrasco}, {Ford}, {Grillo}, {Infante}, {Kelson}, {Lahav}, {Massey}, {Moustakas}, {Rasia}, {Rhodes}, {Vega}, \& {Zheng}}]{Merten_2015}
{Merten}, J., {Meneghetti}, M., {Postman}, M., {et~al.} 2015, \apj, 806, 4

\bibitem[{{Moresco} {et~al.}(2022){Moresco}, {Amati}, {Amendola}, {Birrer}, {Blakeslee}, {Cantiello}, {Cimatti}, {Darling}, {Della Valle}, {Fishbach}, {Grillo}, {Hamaus}, {Holz}, {Izzo}, {Jimenez}, {Lusso}, {Meneghetti}, {Piedipalumbo}, {Pisani}, {Pourtsidou}, {Pozzetti}, {Quartin}, {Risaliti}, {Rosati}, \& {Verde}}]{Moresco_2022}
{Moresco}, M., {Amati}, L., {Amendola}, L., {et~al.} 2022, Living Reviews in Relativity, 25, 6

\bibitem[{{Oguri}(2010)}]{Oguri_2010}
{Oguri}, M. 2010, \pasj, 62, 1017

\bibitem[{{Oke} \& {Gunn}(1983)}]{Oke_1983}
{Oke}, J.~B. \& {Gunn}, J.~E. 1983, \apj, 266, 713

\bibitem[{{Pence} {et~al.}(2010){Pence}, {Chiappetti}, {Page}, {Shaw}, \& {Stobie}}]{pence2010}
{Pence}, W.~D., {Chiappetti}, L., {Page}, C.~G., {Shaw}, R.~A., \& {Stobie}, E. 2010, \aap, 524, A42

\bibitem[{{Pignataro} {et~al.}(2021){Pignataro}, {Bergamini}, {Meneghetti}, {Vanzella}, {Calura}, {Grillo}, {Rosati}, {Angora}, {Brammer}, {Caminha}, {Mercurio}, {Nonino}, \& {Tozzi}}]{Pignataro_2021}
{Pignataro}, G.~V., {Bergamini}, P., {Meneghetti}, M., {et~al.} 2021, \aap, 655, A81

\bibitem[{{Planck Collaboration: Ade} {et~al.}(2016){Planck Collaboration: Ade}, {Aghanim}, {Arnaud}, {Ashdown}, {Aumont}, {Baccigalupi}, {Banday}, {Barreiro}, {Barrena}, {Bartlett}, {Bartolo}, {Battaner}, {Battye}, {Benabed}, {Beno{\^\i}t}, {Benoit-L{\'e}vy}, {Bernard}, {Bersanelli}, {Bielewicz}, {Bikmaev}, {B{\"o}hringer}, {Bonaldi}, {Bonavera}, {Bond}, {Borrill}, {Bouchet}, {Bucher}, {Burenin}, {Burigana}, {Butler}, {Calabrese}, {Cardoso}, {Carvalho}, {Catalano}, {Challinor}, {Chamballu}, {Chary}, {Chiang}, {Chon}, {Christensen}, {Clements}, {Colombi}, {Colombo}, {Combet}, {Comis}, {Couchot}, {Coulais}, {Crill}, {Curto}, {Cuttaia}, {Dahle}, {Danese}, {Davies}, {Davis}, {de Bernardis}, {de Rosa}, {de Zotti}, {Delabrouille}, {D{\'e}sert}, {Dickinson}, {Diego}, {Dolag}, {Dole}, {Donzelli}, {Dor{\'e}}, {Douspis}, {Ducout}, {Dupac}, {Efstathiou}, {Eisenhardt}, {Elsner}, {En{\ss}lin}, {Eriksen}, {Falgarone}, {Fergusson}, {Feroz}, {Ferragamo}, {Finelli}, {Forni}, {Frailis}, {Fraisse}, {Franceschi}, {Frejsel},
  {Galeotta}, {Galli}, {Ganga}, {G{\'e}nova-Santos}, {Giard}, {Giraud-H{\'e}raud}, {Gjerl{\o}w}, {Gonz{\'a}lez-Nuevo}, {G{\'o}rski}, {Grainge}, {Gratton}, {Gregorio}, {Gruppuso}, {Gudmundsson}, {Hansen}, {Hanson}, {Harrison}, {Hempel}, {Henrot-Versill{\'e}}, {Hern{\'a}ndez-Monteagudo}, {Herranz}, {Hildebrandt}, {Hivon}, {Hobson}, {Holmes}, {Hornstrup}, {Hovest}, {Huffenberger}, {Hurier}, {Jaffe}, {Jaffe}, {Jin}, {Jones}, {Juvela}, {Keih{\"a}nen}, {Keskitalo}, {Khamitov}, {Kisner}, {Kneissl}, {Knoche}, {Kunz}, {Kurki-Suonio}, {Lagache}, {Lamarre}, {Lasenby}, {Lattanzi}, {Lawrence}, {Leonardi}, {Lesgourgues}, {Levrier}, {Liguori}, {Lilje}, {Linden-V{\o}rnle}, {L{\'o}pez-Caniego}, {Lubin}, {Mac{\'\i}as-P{\'e}rez}, {Maggio}, {Maino}, {Mak}, {Mandolesi}, {Mangilli}, {Martin}, {Mart{\'\i}nez-Gonz{\'a}lez}, {Masi}, {Matarrese}, {Mazzotta}, {McGehee}, {Mei}, {Melchiorri}, {Melin}, {Mendes}, {Mennella}, {Migliaccio}, {Mitra}, {Miville-Desch{\^e}nes}, {Moneti}, {Montier}, {Morgante}, {Mortlock}, {Moss}, {Munshi},
  {Murphy}, {Naselsky}, {Nastasi}, {Nati}, {Natoli}, {Netterfield}, {N{\o}rgaard-Nielsen}, {Noviello}, {Novikov}, {Novikov}, {Olamaie}, {Oxborrow}, {Paci}, {Pagano}, {Pajot}, {Paoletti}, {Pasian}, {Patanchon}, {Pearson}, {Perdereau}, {Perotto}, {Perrott}, {Perrotta}, {Pettorino}, {Piacentini}, {Piat}, {Pierpaoli}, {Pietrobon}, {Plaszczynski}, {Pointecouteau}, {Polenta}, {Pratt}, {Pr{\'e}zeau}, {Prunet}, {Puget}, {Rachen}, {Reach}, {Rebolo}, {Reinecke}, {Remazeilles}, {Renault}, {Renzi}, {Ristorcelli}, {Rocha}, {Rosset}, {Rossetti}, {Roudier}, {Rozo}, {Rubi{\~n}o-Mart{\'\i}n}, {Rumsey}, {Rusholme}, {Rykoff}, {Sandri}, {Santos}, {Saunders}, {Savelainen}, {Savini}, {Schammel}, {Scott}, {Seiffert}, {Shellard}, {Shimwell}, {Spencer}, {Stanford}, {Stern}, {Stolyarov}, {Stompor}, {Streblyanska}, {Sudiwala}, {Sunyaev}, {Sutton}, {Suur-Uski}, {Sygnet}, {Tauber}, {Terenzi}, {Toffolatti}, {Tomasi}, {Tramonte}, {Tristram}, {Tucci}, {Tuovinen}, {Umana}, {Valenziano}, {Valiviita}, {Van Tent}, {Vielva}, {Villa}, {Wade},
  {Wandelt}, {Wehus}, {White}, {Wright}, {Yvon}, {Zacchei}, \& {Zonca}}]{Ade_2016}
{Planck Collaboration: Ade}, P.~A.~R., {Aghanim}, N., {Arnaud}, M., {et~al.} 2016, \aap, 594, A27

\bibitem[{{Postman} {et~al.}(2012){Postman}, {Coe}, {Ben{\'{\i}}tez}, {Bradley}, {Broadhurst}, {Donahue}, {Ford}, {Graur}, {Graves}, {Jouvel}, {Koekemoer}, {Lemze}, {Medezinski}, {Molino}, {Moustakas}, {Ogaz}, {Riess}, {Rodney}, {Rosati}, {Umetsu}, {Zheng}, {Zitrin}, {Bartelmann}, {Bouwens}, {Czakon}, {Golwala}, {Host}, {Infante}, {Jha}, {Jimenez-Teja}, {Kelson}, {Lahav}, {Lazkoz}, {Maoz}, {McCully}, {Melchior}, {Meneghetti}, {Merten}, {Moustakas}, {Nonino}, {Patel}, {Reg{\"o}s}, {Sayers}, {Seitz}, \& {Van der Wel}}]{Postman_2012_clash}
{Postman}, M., {Coe}, D., {Ben{\'{\i}}tez}, N., {et~al.} 2012, \apjs, 199, 25

\bibitem[{{Price-Whelan} {et~al.}(2018){Price-Whelan}, {Sip{\H{o}}cz}, {G{\"u}nther}, {Lim}, {Crawford}, {Conseil}, {Shupe}, {Craig}, {Dencheva}, {Ginsburg}, {VanderPlas}, {Bradley}, {P{\'e}rez-Su{\'a}rez}, {de Val-Borro}, {Paper Contributors}, {Aldcroft}, {Cruz}, {Robitaille}, {Tollerud}, {Coordination Committee}, {Ardelean}, {Babej}, {Bach}, {Bachetti}, {Bakanov}, {Bamford}, {Barentsen}, {Barmby}, {Baumbach}, {Berry}, {Biscani}, {Boquien}, {Bostroem}, {Bouma}, {Brammer}, {Bray}, {Breytenbach}, {Buddelmeijer}, {Burke}, {Calderone}, {Cano Rodr{\'\i}guez}, {Cara}, {Cardoso}, {Cheedella}, {Copin}, {Corrales}, {Crichton}, {D{\textquoteright}Avella}, {Deil}, {Depagne}, {Dietrich}, {Donath}, {Droettboom}, {Earl}, {Erben}, {Fabbro}, {Ferreira}, {Finethy}, {Fox}, {Garrison}, {Gibbons}, {Goldstein}, {Gommers}, {Greco}, {Greenfield}, {Groener}, {Grollier}, {Hagen}, {Hirst}, {Homeier}, {Horton}, {Hosseinzadeh}, {Hu}, {Hunkeler}, {Ivezi{\'c}}, {Jain}, {Jenness}, {Kanarek}, {Kendrew}, {Kern}, {Kerzendorf}, {Khvalko},
  {King}, {Kirkby}, {Kulkarni}, {Kumar}, {Lee}, {Lenz}, {Littlefair}, {Ma}, {Macleod}, {Mastropietro}, {McCully}, {Montagnac}, {Morris}, {Mueller}, {Mumford}, {Muna}, {Murphy}, {Nelson}, {Nguyen}, {Ninan}, {N{\"o}the}, {Ogaz}, {Oh}, {Parejko}, {Parley}, {Pascual}, {Patil}, {Patil}, {Plunkett}, {Prochaska}, {Rastogi}, {Reddy Janga}, {Sabater}, {Sakurikar}, {Seifert}, {Sherbert}, {Sherwood-Taylor}, {Shih}, {Sick}, {Silbiger}, {Singanamalla}, {Singer}, {Sladen}, {Sooley}, {Sornarajah}, {Streicher}, {Teuben}, {Thomas}, {Tremblay}, {Turner}, {Terr{\'o}n}, {van Kerkwijk}, {de la Vega}, {Watkins}, {Weaver}, {Whitmore}, {Woillez}, {Zabalza}, \& {Contributors}}]{astropy2}
{Price-Whelan}, A.~M., {Sip{\H{o}}cz}, B.~M., {G{\"u}nther}, H.~M., {et~al.} 2018, \aj, 156, 123

\bibitem[{{Sartoris} {et~al.}(2016){Sartoris}, {Biviano}, {Fedeli}, {Bartlett}, {Borgani}, {Costanzi}, {Giocoli}, {Moscardini}, {Weller}, {Ascaso}, {Bardelli}, {Maurogordato}, \& {Viana}}]{Sartoris_2016}
{Sartoris}, B., {Biviano}, A., {Fedeli}, C., {et~al.} 2016, \mnras, 459, 1764

\bibitem[{{Treu} {et~al.}(2022){Treu}, {Suyu}, \& {Marshall}}]{Treu_2022}
{Treu}, T., {Suyu}, S.~H., \& {Marshall}, P.~J. 2022, \aapr, 30, 8

\bibitem[{{van der Walt} {et~al.}(2011){van der Walt}, {Colbert}, \& {Varoquaux}}]{numpy1}
{van der Walt}, S., {Colbert}, S.~C., \& {Varoquaux}, G. 2011, Computing in Science Engineering, 13, 22

\bibitem[{Van~Rossum \& Drake(2009)}]{python}
Van~Rossum, G. \& Drake, F.~L. 2009, Python 3 Reference Manual (Scotts Valley, CA: CreateSpace)

\bibitem[{{Virtanen} {et~al.}(2020){Virtanen}, {Gommers}, {Oliphant}, {Haberland}, {Reddy}, {Cournapeau}, {Burovski}, {Peterson}, {Weckesser}, {Bright}, {van der Walt}, {Brett}, {Wilson}, {Jarrod Millman}, {Mayorov}, {Nelson}, {Jones}, {Kern}, {Larson}, {Carey}, {Polat}, {Feng}, {Moore}, {Vand erPlas}, {Laxalde}, {Perktold}, {Cimrman}, {Henriksen}, {Quintero}, {Harris}, {Archibald}, {Ribeiro}, {Pedregosa}, {van Mulbregt}, \& {Contributors}}]{scipy}
{Virtanen}, P., {Gommers}, R., {Oliphant}, T.~E., {et~al.} 2020, Nature Methods, 17, 261

\bibitem[{{Wisotzki} {et~al.}(2002){Wisotzki}, {Schechter}, {Bradt}, {Heinm{\"u}ller}, \& {Reimers}}]{Wisotzki_2002}
{Wisotzki}, L., {Schechter}, P.~L., {Bradt}, H.~V., {Heinm{\"u}ller}, J., \& {Reimers}, D. 2002, \aap, 395, 17

\bibitem[{{Zitrin} \& {Broadhurst}(2009)}]{Zitrin_2009}
{Zitrin}, A. \& {Broadhurst}, T. 2009, \apjl, 703, L132

\bibitem[{{Zitrin} {et~al.}(2013){Zitrin}, {Meneghetti}, {Umetsu}, {Broadhurst}, {Bartelmann}, {Bouwens}, {Bradley}, {Carrasco}, {Coe}, {Ford}, {Kelson}, {Koekemoer}, {Medezinski}, {Moustakas}, {Moustakas}, {Nonino}, {Postman}, {Rosati}, {Seidel}, {Seitz}, {Sendra}, {Shu}, {Vega}, \& {Zheng}}]{Zitrin_2013}
{Zitrin}, A., {Meneghetti}, M., {Umetsu}, K., {et~al.} 2013, \apjl, 762, L30

\end{thebibliography}

\label{LastPage}

\end{document}